\documentclass[journal,draftcls,onecolumn,12pt,twoside]{IEEEtran}
\usepackage{graphicx}
\usepackage{amsmath, amsfonts, amssymb, amsthm}
\usepackage{enumerate}
\usepackage{multirow}
\usepackage{color}
\usepackage{longtable}

\theoremstyle{definition} \newtheorem{theorem}{Theorem}
\theoremstyle{definition} \newtheorem{corollary}[theorem]{Corollary}
\theoremstyle{definition} \newtheorem{proposition}[theorem]{Proposition}
\theoremstyle{definition} \newtheorem{definition}[theorem]{Definition}
\theoremstyle{definition} \newtheorem{lemma}[theorem]{Lemma}
\theoremstyle{definition} 
\theoremstyle{definition} 
\theoremstyle{definition} 
\theoremstyle{definition} \newtheorem*{notations}{Notation}
\theoremstyle{definition} \newtheorem*{example}{Example}
{\end{list}}

\begin{document}
\title{Circular-shift Linear Network Coding}
\author{\IEEEauthorblockN{Hanqi~Tang\textsuperscript{\dag},~Qifu~Tyler~Sun\textsuperscript{\dag},~Zongpeng~Li\textsuperscript{\ddag{*}},~Xiaolong Yang\textsuperscript{\dag},~and~Keping~Long\textsuperscript{\dag}} \\
\IEEEauthorblockA{\textsuperscript{\dag}University of Science and Technology Beijing, P. R. China \\
\textsuperscript{\ddag}University of Calgary, Canada~~~
\textsuperscript{*} Wuhan University, P. R. China\\
}
\thanks{$^\dag~$Q. T. Sun (Email: qfsun@ustb.edu.cn) is the corresponding author.}%\newline \indent This work was partially supported by the National Natural Science Foundation of China (No. 61471034 \& 61571335) and the National Science and Engineering Research Council of Canada (NSERC No. RT733206).}
}% <-this % stops a space
\maketitle
\sloppy

\begin{abstract}
We study a class of linear network coding (LNC) schemes, called \emph{circular-shift} LNC, whose encoding operations consist of only circular-shifts and bit-wise additions (XOR). %
Formulated as a special vector linear code over GF($2$), an $L$-dimensional circular-shift linear code of degree $\delta$ restricts its local encoding kernels to be the summation of at most $\delta$ cyclic permutation matrices of size $L$. %
We show that on a general network, for a certain block length $L$, every scalar linear solution over GF($2^{L-1}$) can induce an $L$-dimensional circular-shift linear solution with 1-bit redundancy per-edge transmission. %
Consequently, specific to a multicast network, such a circular-shift linear solution of an arbitrary degree $\delta$ can be efficiently constructed, which has an interesting complexity tradeoff between encoding and decoding with different choices of $\delta$. %
By further proving that circular-shift LNC is insufficient to achieve the exact capacity of certain multicast networks, we show the optimality of the efficiently constructed circular-shift linear solution in the sense that its 1-bit redundancy is inevitable. %
Finally, both theoretical and numerical analysis imply that with increasing $L$, a randomly constructed circular-shift linear code has linear solvability behavior comparable to a randomly constructed permutation-based linear code, but has shorter overheads.
\end{abstract}

\section{Introduction}
Assume that every edge in a network transmits a binary sequence of length $L$. Different linear network coding (LNC) schemes manipulate the binary sequences by different approaches. With conventional scalar LNC (See, e.g., \cite{LiYeungCai03}\cite{KoetterMedard03}) and vector LNC (See, e.g., \cite{Medard03}\cite{Ebrahimi}), the binary sequence carried at every edge is modeled, respectively, as an element of the finite field GF($2^L$) and an $L$-dimensional vector over GF($2$). The coding operations performed at every intermediate node by scalar LNC and by vector LNC are linear functions over GF($2^L$) and over the ring of $L \times L$ binary matrices, respectively. %
The coefficients of these linear functions are called the local encoding kernels (See, e.g., \cite{Yeung_ITNC_Book}\cite{Sun_TCom16}).

There have been continuous attempts to design LNC schemes with low implementation complexities. A straightforward way is to reduce the block length $L$. It is well known that when $2^L$ is no smaller than the number of receivers, a scalar linear solution over GF($2^L$) can be efficiently constructed on a (single-source) multicast network by algorithms in \cite{Jaggi05} and \cite{Langberg09}. Recent literature has witnessed a few interesting multicast networks that have an $L$-dimensional vector linear solution over GF($2$) but do not have a scalar linear solution over GF($2^{L'}$) for any $L' \leq L$  \cite{Sun_TCom16}\cite{Etzion16}. In particular, for the multicast networks designed in \cite{Etzion16}, the minimum block length $L$ for an $L$-dimensional vector linear solution over GF($2$) can be much shorter than the minimum block length $L'$ for a scalar linear solution over GF($2^{L'}$). This verifies that compared with scalar LNC, vector LNC may yield solutions with lower implementation complexities.

Another approach to reduce the encoding complexity of LNC is to carefully design the coding operations performed at intermediate nodes. A special type of vector LNC based on permutation operations is studied in \cite{Jaggi06}, from a random coding approach. In permutation-based vector LNC, at an intermediate node, every incoming binary sequence is first permuted, and then an outgoing binary sequence is formed by bit-wise additions of the permutated incoming binary sequences. Equivalently, the local encoding kernels at intermediate nodes are chosen from $L \times L$ binary permutation matrices, rather than arbitrary $L \times L$ binary matrices. Though permutation can be more efficiently implemented than general matrix multiplication on a binary sequence, its computational complexity may not be low enough for real-world implementation, when the block length $L$ is long, as required in random coding.

Towards further reducing the encoding and decoding complexity of LNC, %

we study in this paper another class of LNC schemes whose encoding operations on the binary sequences are restricted to merely bit-wise additions and \emph{circular-shifts}, which are operations to sequentially move the final entry to the first position, and shift all other entries to the next position.
Circular-shift operations have lower computational complexity than permutations, and are amenable to implementation through atomic hardware operations. %

One may notice that prior to this work, similar ideas of adopting circular-shift and bit-wise addition operations for encoding have been considered in \cite{Xiao07}, \cite{Khojastepour10} and \cite{HouShum16}. In particular, the LNC schemes studied in \cite{Xiao07}, for a special class of multicast networks called Combination Networks, involve not only circular-shifts and bit-wise additions, but also a bit truncation process. The low-complexity LNC schemes studied in \cite{Khojastepour10}, for an arbitrary multicast network, are called rotation-and-add linear codes, and the low-complexity functional-repair regenerating codes studied in \cite{HouShum16} for a distributed system are called BASIC (Binary Addition and Shift Implementable Cyclic convolutional) functional-repair regenerating codes. From the perspective of cyclic convolutional coding, the work in \cite{Khojastepour10} and \cite{HouShum16} respectively showed the \emph{existence} of the rotation-and-add linear solutions and BASIC functional-repair regenerating codes. %
However, due to the lack of a systematic model, they did not provide any efficient algorithm to construct these codes and how to decode these codes was not discussed either.

In this paper, we algebraically formulate circular-shift LNC as a special type of vector LNC. %

In particular, an $L$-dimensional circular-shift linear code of degree $\delta$ is defined as an $L$-dimensional vector linear code over GF($2$) with the local encoding kernels restricted to the summation of at most $\delta$ cyclic permutation matrices of size $L$. %
Under this framework, we make the following contributions for the theory of circular-shift LNC:
\begin{itemize}
\item An intrinsic connection between scalar LNC and circular-shift LNC is established on a \emph{general multi-source multicast network}. %
    In particular, for a prime $L$ with primitive root $2$, \emph{i.e.}, with the multiplicative order of $2$ modulo $L$ equal to $L-1$, every scalar linear solution over GF($2^{L-1}$) can induce an $(L-1, L)$ circular-shift linear solution of degree at most $\frac{L-1}{2}$. The notation $(L-1, L)$ here means that for this $L$-dimensional circular-shift linear code, the binary sequences generated at sources and transmitted along edges are respectively of lengths $L-1$ and $L$, so that the induced code falls into the category of \emph{fractional} LNC (See, e.g., \cite{Zeger16}). %
\item Consequently, specific to a (single-source) multicast network, an $(L-1, L)$ circular-shift linear solution of an \emph{arbitrary} degree $\delta$ can be efficiently constructed. In addition, we analyze that when $\delta = \frac{L-1}{2}$, the constructed solution requires fewer binary operations for both encoding and decoding processes compared with scalar linear solutions over GF($2^{L-1}$). Furthermore, when $\delta$ decreases from $\frac{L-1}{2}$ to $1$, there is an interesting tradeoff between decreasing encoding complexity and increasing decoding complexity, making the code design more flexible.
\item We further prove that circular-shift LNC is insufficient to achieve the exact capacity of certain multicast networks. This result in turn shows the optimality of the efficiently constructed circular-shift linear solution for a multicast network in the sense that the 1-bit redundancy of the code is inevitable.
\item We also study circular-shift LNC from a random coding approach. We derive a lower bound on the success probability of randomly generating a circular-shift linear solution, which is essentially the same as the one in \cite{Jaggi06} for permutation-based LNC. Numerical results also demonstrate comparable success probability of randomly generating a circular-shift linear solution to the one of randomly generating a permutation-based linear solution. These findings are interesting because for a block length $L$, circular-shift LNC can only provide $L+1$ local encoding kernel candidates, much less than $L!$ in permutation-based LNC. Last, we show that circular-shift LNC has the additional advantage of shorter overheads for random coding.
\end{itemize}

Because both the rotation-and-add linear codes (over GF(2)) and the BASIC functional-repair regenerating codes can be regarded as circular-shift linear codes of degree $1$, the present paper also unveils a method to efficiently construct these codes.

The rest of the paper is organized as follows. %
Section II briefly reviews the basic concepts of LNC as well as some useful properties of cyclic permutation matrices. %
Section III formulates circular-shift LNC from the perspective of vector LNC and establishes an intrinsic connection between scalar LNC and circular-shift LNC on general networks. %
Section IV discusses efficient construction of circular-shift linear solutions on multicast networks. %
Section V analyzes circular-shift LNC by the random coding approach. %
Section VII concludes the paper. %

In addition to the proof details of some lemmas and propositions, the frequently used important notation for the discussion of circular-shift LNC is listed in Appendix for reference.

\section{Preliminaries}
\subsection{Linear Network Codes}
A general (acyclic multi-source multicast) network is modeled as a finite directed acyclic multigraph, with a set $S$ of source nodes and a set $T$ of receivers.
For a node $v$ in the network, denote by $\mathrm{In}(v)$ and $\mathrm{Out}(v)$, respectively, the set of its incoming and outgoing edges. Similarly, for a set $N$ of nodes, denote by $\mathrm{In}(N)$ and $\mathrm{Out}(N)$ the set of incoming edges to and outgoing edges from the nodes in $N$, \emph{i.e.}, $\mathrm{In}(N) = \bigcup_{v\in N}\mathrm{In}(v)$ and $\bigcup_{v\in N}\mathrm{Out}(v)$.
Every edge has a unit capacity to transmit a data unit per channel use. Write $|\mathrm{Out}(S)| = \omega$. Every source $s \in S$ generates $|\mathrm{Out}(s)|$ source data units, and there are in total $\omega$ source data units generated by $S$ to be propagated along the network.  Assume an arbitrary order on $S = \{s_1, \ldots, s_{|S|}\}$ and a topological order on the edge set $E$ of the network led by the edges in $\mathrm{Out}(s_j)$, $1 \leq j\leq |S|$, sequentially. For every receiver $t \in T$, based on the data units received from edges in $\mathrm{In}(t)$, its goal is to recover the $\omega_t = |\mathrm{Out}(S_t)|$ data units generated from a particular set $S_t \subseteq S$ of sources. %
To simplify the network model, without loss of generality (WLOG), assume that for every source, its in-degree is zero and there is not any edge leading from it to a receiver.
When there is a unique source node $s$ and all receivers need recover the $\omega$ source data units generated at $s$, the network is called a \emph{multicast network}. In a multicast network, the maximum flow from the source to every receiver is assumed equal to $\omega$.

\begin{notations}
Let $\otimes$ denote the Kronecker product and $\mathbf{u}_e$ be an $\omega \times 1$ unit vector such that the column-wise juxtaposition\footnote{Unless otherwise specified, all juxtaposition of matrices or vectors throughout this paper refers to column-wise juxtaposition.} $[\mathbf{u}_e]_{e \in \mathrm{Out}(S)}$ forms the $\omega \times \omega$ identity matrix $\mathbf{I}_\omega$. For a positive integer $j$, define $\mathbf{U}_e^j = \mathbf{u}_e\otimes \mathbf{I}_j$. Note that $\mathbf{U}_e^j $ is an $\omega j \times j$ matrix and $[\mathbf{U}_e^j]_{e \in \mathrm{Out}(S)} = \mathbf{I}_{\omega j}$.
\end{notations}

For vector LNC, the data unit transmitted along every edge $e$ is an $L$-dimensional \emph{row vector} $\mathbf{m}_e$ of binary data symbols. An \emph{$L$-dimensional vector linear code $(\mathbf{K}_{d,e})$ over} GF(2) (See, e.g., \cite{Sun_TCom16}), is an assignment of a local encoding kernel $\mathbf{K}_{d,e}$, which is an $L\times L$ matrix over GF(2), to every pair $(d, e)$ of edges such that $\mathbf{K}_{d,e}$ is the zero matrix $\mathbf{0}$ when $(d, e)$ is not an adjacent pair. Then, for every edge $e$ emanating from a non-source node $v$, the data unit vector of binary data symbols transmitted on $e$ is $\mathbf{m}_e = \sum_{d\in \mathrm{In}(v)} \mathbf{m}_d\mathbf{K}_{d,e}$. WLOG, for every $s \in S$, assume the data units $\mathbf{m}_e$, $e \in \mathrm{Out}(s)$, just constitute the $|\mathrm{Out}(s)|$ source data units generated by $s$.
Every vector linear code uniquely determines a global encoding kernel $\mathbf{F}_{e}$, which is an $\omega L\times L$ matrix over GF(2), for every edge $e$ such that
\begin{itemize}
\item $[\mathbf{F}_e]_{e\in \mathrm{Out}(S)} = [\mathbf{U}_e^L]_{e\in \mathrm{Out}(S)} = \mathbf{I}_{\omega L}$;
\vspace{3pt}
\item For every outgoing edge $e$ from a non-source node $v$, $\mathbf{F}_e = \sum_{d\in \mathrm{In}(v)} \mathbf{F}_d\mathbf{K}_{d,e}$.
\end{itemize} %
Correspondingly, the data unit vector transmitted along every edge can also be represented as
\begin{equation}
\label{eqn:def_m_e}
\mathbf{m}_e = [\mathbf{m_d}]_{d\in \mathrm{Out}(S)} \mathbf{F}_e.
\end{equation}
A vector linear code is called a \emph{vector linear solution} if for every receiver $t \in T$, there is an $|\mathrm{In}(t)|L\times \omega_t L$ decoding matrix $\mathbf{D}_t$ over GF(2) such that
\begin{equation}
\label{eqn:decoding_kernel}
[\mathbf{F}_e]_{e\in \mathrm{In}(t)}\mathbf{D}_t = [\mathbf{U}_e^L]_{e\in \mathrm{Out}(S_t)}
\end{equation}
Based on $\mathbf{D}_t$, the data units generated at sources in $S_t$ can be recovered by receiver $t$ via

\begin{align}
\label{eqn:decoding_symbol}
[\mathbf{m}_e]_{e\in \mathrm{In}(t)}\mathbf{D}_t &= \left([\mathbf{m_d}]_{d\in \mathrm{Out}(S)} [\mathbf{F}_e]_{e\in \mathrm{In}(t)} \right) \mathbf{D}_t \\
&=[\mathbf{m_d}]_{d\in \mathrm{Out}(S)} \left([\mathbf{F}_e]_{e\in \mathrm{In}(t)} \mathbf{D}_t\right) \\
&= [\mathbf{m_d}]_{d\in \mathrm{Out}(S)} [\mathbf{U}_e^L]_{e\in \mathrm{Out}(S_t)} \\
&= [\mathbf{m_d}]_{d\in \mathrm{Out}(S_t)}.
\end{align}

In network coding theory, there are networks, such as the famous V\'amos Network designed in \cite{Zeger07}, with the linear coding capacity equal to a rational number. Thus, in order to achieve the rational linear coding capacity, vector LNC is insufficient and what we need is \emph{fractional LNC}, a generalization of vector LNC (See, e.g., \cite{Zeger16}).
Same as in an $L$-dimensional vector linear code over GF(2), in an $(L', L)$-fractional linear code over GF(2), the data unit $\mathbf{m}_e$ transmitted on every edge $e$ is an $L$-dimensional row vector over GF(2), and the local encoding kernels $\mathbf{K}_{d, e}$ are $L \times L$ matrices over GF(2). The difference is that for an $(L', L)$-fractional linear code, where $L' \leq L$, the $|\mathrm{Out}(s)|$ data units generated at every source $s \in S$ are $L'$-dimensional row vectors over GF(2). By a slight abuse of notation, denote the $|\mathrm{Out}(s)|$ $L'$-dimensional row vectors generated at $s$ by $\mathbf{m}'_e$, $e \in \mathrm{Out}(s)$. Each of the $L$ binary data symbols in the data unit $\mathbf{m}_e$ transmitted on $e \in \mathrm{Out}(s)$, is a GF(2)-linear combination of the ones in $\mathbf{m}'_e$, $e \in \mathrm{Out}(s)$, \emph{i.e.},
\begin{equation}
[\mathbf{m}_e]_{e \in \mathrm{Out}(s)} = [\mathbf{m}_e']_{e \in \mathrm{Out}(s)}\mathbf{G}_s
\end{equation}
for some $|\mathrm{Out}(s)|L' \times |\mathrm{Out}(s)| L$ matrix $\mathbf{G}_s$ over GF(2). %
In total, the data units $\mathbf{m}_e$ transmitted on $e \in \mathrm{Out}(S)$ can be expressed as
\begin{equation}
[\mathbf{m}_e]_{e\in \mathrm{Out}(S)} = [\mathbf{m}_e']_{e\in \mathrm{Out}(S)} \mathbf{G}_S, %= [\mathbf{m}_e']_{e\in Out(S)} \mathcal{D}(\mathbf{G}_s)_{s \in S},
\end{equation}
where $\mathbf{G}_S$ denotes the $\omega L' \times \omega L$ matrix
\begin{equation}
\setlength{\arraycolsep}{3.0pt}
\renewcommand{\arraystretch}{0.8}
\mathbf{G}_{S}= \left[\begin{matrix}
\mathbf{G}_{s_1} & \mathbf{0} & \ldots & \mathbf{0}  \\
\mathbf{0} & \mathbf{G}_{s_2} & \ldots & \mathbf{0} \\
\vdots & \vdots & \ddots & \vdots  \\
\mathbf{0} & \mathbf{0} & \mathbf{0} & \mathbf{G}_{s_{|S|}}
\end{matrix}\right]
\end{equation}
which consists of $|S|\times |S|$ blocks with the $(j, j)^{th}$ ``diagonal'' block, $1 \leq j \leq |S|$, being the $\omega_{s_j}L' \times \omega_{s_j}L$ matrix $\mathbf{G}_{s_j}$.

Therefore, an $(L', L)$-fractional linear code $(\mathbf{K}_{d,e})$ over GF($2$) is an $L$-dimensional vector linear code $(\mathbf{K}_{d,e})$ over GF($2$) with an additional $\omega_{s_j}L' \times \omega_{s_j}L$ binary matrix $\mathbf{G}_{s_j}$ for every source $s_j$. It qualifies as an $(L', L)$-\emph{fractional linear solution} if for each receiver $t$, there is an $|\mathrm{In}(t)| L \times \omega_t L'$ matrix $\mathbf{D}_t$ over GF(2) such that
\begin{equation}
\mathbf{G}_S[\mathbf{F}_e]_{e\in \mathrm{In}(t)}\mathbf{D}_t = [\mathbf{U}_e^{L'}]_{e \in \mathrm{Out}(S_t)}.
\end{equation}
Based on the decoding matrix $\mathbf{D}_t$, the data units $\mathbf{m}'_e$, $e \in \mathrm{Out}(S_t)$ generated by sources in $S_t$ can be recovered at $t$ via
\begin{align}
[\mathbf{m}_e]_{e\in \mathrm{In}(t)}\mathbf{D}_t &= \left([\mathbf{m}_e]_{e\in \mathrm{Out}(S)}[\mathbf{F}_e]_{e\in \mathrm{In}(t)}\right)\mathbf{D}_t \\
&= \left([\mathbf{m}_e']_{e\in \mathrm{Out}(S)} \mathbf{G}_S[\mathbf{F}_e]_{e\in \mathrm{In}(t)}\right)\mathbf{D}_t \\
&= [\mathbf{m}_e']_{e\in \mathrm{Out}(S)}[\mathbf{U}_e^{L'}]_{e \in \mathrm{Out}(S_t)} \\
&= [\mathbf{m}_e']_{e \in \mathrm{Out}(S_t)}.
\end{align}

Conventional scalar linear codes over GF(2) and $L$-dimensional vector linear codes over GF(2) can be respectively regarded as $(1, 1)$-fractional and $(L, L)$-fractional linear codes over GF(2), with the matrix $\mathbf{G}_{s_j}$ for every source $s_j$ equal to the identity matrix $\mathbf{I}_{\omega_{s_j}L}$. In a scalar linear code over GF($2^L$), instead of $\mathbf{K}_{d,e}$ and $\mathbf{F}_{e}$, we shall use the scalar symbol $k_{d,e}$ and the vector symbol $\mathbf{f}_e$ to denote the local encoding kernels and global encoding kernels respectively.

\begin{figure}[!t]
\centering
\scalebox{0.78}
{\includegraphics{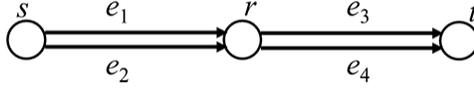}}

\caption{A network consists of three nodes.}

\label{Fig:3-node-newtork}
\end{figure}

\begin{example}
Consider the network depicted in Fig.\ref{Fig:3-node-newtork}, which consists of a source node $s$, a relay node $r$ and a receiver $t$. Every edge can transmit a binary sequence of length $3$. Source $s$ generates two binary sequences $(m_{11}, m_{12})$, $(m_{21}, m_{22})$ of length $2$. Consider a $(2, 3)$-fractional linear code over GF($2$) with the $4\times 6$ encoding matrix $\mathbf{G}_s$ at $s$ to be $\setlength{\arraycolsep}{2.0pt}
\renewcommand{\arraystretch}{0.5}
\mathbf{G}_s = \begin{bmatrix}
1 & 0 & 0 & 0 & 0 & 0 \\
0 & 0 & 0 & 1 & 0 & 0 \\
0 & 0 & 1 & 0 & 0 & 0 \\
0 & 0 & 0 & 0 & 0 & 1
\end{bmatrix}$, %
and the local encoding kernels at $r$ to be $\mathbf{K}_{e_1,e_3} = \setlength{\arraycolsep}{2.0pt}
\renewcommand{\arraystretch}{0.5}
\begin{bmatrix}
0 & 1 & 0  \\
0 & 0 & 0  \\
0 & 1 & 1
\end{bmatrix}$, %
$\setlength{\arraycolsep}{2.0pt}
\renewcommand{\arraystretch}{0.5}
\mathbf{K}_{e_1,e_4} = \begin{bmatrix}
0 & 0 & 0  \\
0 & 0 & 0  \\
0 & 0 & 0
\end{bmatrix}$, %
$\setlength{\arraycolsep}{2.0pt}
\renewcommand{\arraystretch}{0.5}
\mathbf{K}_{e_2,e_3} = \begin{bmatrix}
0 & 0 & 0  \\
0 & 0 & 0  \\
0 & 0 & 0
\end{bmatrix}$ %
$\setlength{\arraycolsep}{2.0pt}
\renewcommand{\arraystretch}{0.5}
\mathbf{K}_{e_2,e_4} = \begin{bmatrix}
0 & 1 & 0  \\
0 & 0 & 0  \\
0 & 1 & 1
\end{bmatrix}$. Under this code, the data units $\mathbf{m}_{e_j}$ transmitted on edges $e_j$, $1 \leq j \leq 4$ are
$\mathbf{m}_{e_1} = [m_{11}~0~m_{21}],~\mathbf{m}_{e_2} = [m_{12}~0~m_{22}],~\mathbf{m}_{e_3} = [0~m_{11}+m_{21}~m_{21}],~\mathbf{m}_{e_4} = [0~m_{12}+m_{22}~m_{22}]$. %
Correspondingly, the juxtaposition of global encoding kernels for edges incoming to $t$ are %
\begin{equation}
\setlength{\arraycolsep}{3.0pt}
\renewcommand{\arraystretch}{0.7}
[\mathbf{F}_e]_{e\in \mathrm{In}(t)} = [\mathbf{F}_{e_3}~\mathbf{F}_{e_4}] =
\left[
\begin{matrix}
\mathbf{K}_{e_1,e_3} & \mathbf{K}_{e_1,e_4}  \\
\mathbf{K}_{e_2,e_3} & \mathbf{K}_{e_2,e_4}  \\
\end{matrix}
\right]
=
\setlength{\arraycolsep}{3.0pt}
\renewcommand{\arraystretch}{0.6}
\begin{bmatrix}
0 & 1 & 0 & 0 & 0 & 0 \\
0 & 0 & 0 & 0 & 0 & 0 \\
0 & 1 & 1 & 0 & 0 & 0 \\
0 & 0 & 0 & 0 & 1 & 0 \\
0 & 0 & 0 & 0 & 0 & 0 \\
0 & 1 & 0 & 0 & 1 & 1
\end{bmatrix}
\end{equation}
Given the $6\times 4$ matrix $\setlength{\arraycolsep}{2.0pt}
\renewcommand{\arraystretch}{0.5}
\mathbf{D}_t = \begin{bmatrix}
0 & 0 & 0 & 0 \\
1 & 0 & 0 & 0 \\
1 & 0 & 1 & 0 \\
0 & 0 & 0 & 0 \\
0 & 1 & 0 & 0 \\
0 & 1 & 0 & 1
\end{bmatrix}$, as $\mathbf{G}_s[\mathbf{F}_e]_{e\in \mathrm{In}(t)}\mathbf{D}_t = \mathbf{I}_4$, $\mathbf{D}_t$ is the decoding matrix for receiver $t$, which can recover the source data units via $[\mathbf{m}_{e_3}~\mathbf{m}_{e_4}]\mathbf{D}_t = [m_{11}~m_{12}~m_{21}~m_{22}]$. The considered code is thus a $(2,3)$-fractional linear solution.
\end{example}

\subsection{Cyclic Permutation Matrices}
For a positive integer $L$, denote by $\mathbf{C}_L$ the following $L \times L$ \emph{cyclic permutation matrix} (over GF($2$))
\begin{equation}
\label{eqn:cyclic_permutation_matrix}
\setlength{\arraycolsep}{3.0pt}
\renewcommand{\arraystretch}{0.7}
\mathbf{C}_L = \left[\begin{matrix}
0 & 1 & 0 & \ldots & 0 \\
0 & 0 & 1 & \ddots & 0 \\
0 & \ddots & \ddots & \ddots & 0 \\
0 & \ddots & \ddots & 0 & 1 \\
1 & 0 & \ldots & 0 & 0 \end{matrix}\right].
\end{equation}
For a binary row vector $\mathbf{m} = [m_1~m_2~\ldots~ m_L]$, the linear operation $\mathbf{m}\mathbf{C}^j$ is equivalent to a circular-shift of $\mathbf{m}$ by $j$ bits to the right, that is, $\forall~0 < j < L$,
\begin{equation}
[m_1~ m_2~ \ldots~ m_L]\mathbf{C}^j = [m_{L-j+1}~\ldots~m_L~ m_1~ \ldots~ m_{L-j}].
\end{equation}

The following diagonalization manipulation on $\mathbf{C}_L$ over a larger field will be very useful for our subsequent study of circular-shift LNC in Section III.

\begin{lemma}
\label{lemma:cyclic_matrix_decomposition}
Let $L$ be an odd integer and $\alpha$ be a primitive $L^{th}$ root of unity over GF($2$). Denote by $\mathbf{V}_L$ the $L\times L$ Vandermonde matrix generated by $1, \alpha, \ldots, \alpha^{L-1}$ over GF(2)($\alpha$), the minimal field containing GF($2$) and $\alpha$:
\begin{equation}
\label{eqn:Vandermonde_p}
\setlength{\arraycolsep}{3.0pt}
\renewcommand{\arraystretch}{0.8}
\mathbf{V}_L = \left[
\begin{matrix}
1 & 1 & \ldots & 1 \\
1 & \alpha & \ldots & \alpha^{L-1} \\
\vdots & \vdots & \ldots & \vdots \\
1 & \alpha^{L-1} & \ldots & \alpha^{(L-1)(L-1)}
\end{matrix}
\right],
\end{equation}
and by $\mathbf{\Lambda}_\alpha$ the $L \times L$ diagonal matrix with diagonal entries equal to $1, \alpha, \ldots, \alpha^{L-1}$, {\em i.e.},
\begin{equation}
\setlength{\arraycolsep}{3.0pt}
\renewcommand{\arraystretch}{0.7}
\mathbf{\Lambda}_\alpha =
\left[\begin{matrix}
1 & 0 & \ldots & 0 \\
0 & \alpha & \ddots & \vdots \\
\vdots & \ddots & \ddots & 0 \\
0 & \ldots & 0 & \alpha^{L-1}
\end{matrix}\right].
\end{equation}
The inverse of $\mathbf{V}_L$ is
\begin{equation}
\label{eqn:Vandermond_L_inverse}
\setlength{\arraycolsep}{3.0pt}
\renewcommand{\arraystretch}{0.7}
\mathbf{V}_L^{-1} =
\left[
\begin{matrix}
1 & 1 & \ldots & 1 \\
1 & \alpha^{-1} & \ldots & \alpha^{-(L-1)} \\
\vdots & \vdots & \ldots & \vdots \\
1 & \alpha^{-(L-1)} & \ldots & \alpha^{-(L-1)(L-1)}
\end{matrix}
\right],
\end{equation}
and
\begin{equation}
\label{eqn:cyclic_matrix_decomposition}
\mathbf{C}_L^i = \mathbf{V}_L \mathbf{\Lambda}_\alpha^i  \mathbf{V}_L^{-1} ~~~~~ \forall i \geq 0.
\end{equation}
\begin{proof}
It can be proved in a similar way to show Lemma 1 in \cite{HuangQin_TIT_12}. We provide the proof in Appendix-\ref{Appendix:Lemma_cyclic_matrix_decomposition_proof} to make it self-contained.
\end{proof}
\end{lemma}

It is interesting to note that the diagonalization manipulation on $\mathbf{C}_L$ in Lemma \ref{lemma:cyclic_matrix_decomposition} has already been used in the rank analysis of quasi-cyclic LDPC codes \cite{HuangQin_TCom_10}\cite{HuangQin_TIT_12} as well as certain quasi-cyclic stabilizer quantum LDPC codes \cite{XieShane_TCom_18}. The present paper will be its first usage in the construction of linear network codes.

For $1 \leq \delta \leq L$, let $\mathcal{C}_\delta$ denote the following set of matrices:
\begin{equation}
\label{eqn:set_C_delta}
\mathcal{C}_\delta = \left\{\sum\nolimits_{j=0}^{L-1}a_j\mathbf{C}_L^j: a_j \in \{0, 1\}, \sum\nolimits_{j = 0}^{L-1} a_j \leq \delta \right\},
\end{equation}
that is, $\mathcal{C}_\delta$ contains the matrices that are the summation of at most $\delta$ cyclic permutation matrices of size $L$. As a consequence of Lemma \ref{lemma:cyclic_matrix_decomposition}, when $L$ is odd, every matrix $\sum_{j=0}^{L-1}a_j\mathbf{C}_L^j \in \mathcal{C}_\delta$ can be diagonalized as
\begin{equation}
\label{eqn:circulant_matrix_diagonalize}
\sum\nolimits_{j=0}^{L-1}a_j\mathbf{C}_L^j = \mathbf{V}_L\left(\sum\nolimits_{j=0}^{L-1}a_j\mathbf{\Lambda}_\alpha^j\right)\mathbf{V}_L^{-1}. \end{equation}
In addition, since
\begin{equation}
\setlength{\arraycolsep}{3.0pt}
\renewcommand{\arraystretch}{0.7}
\sum\nolimits_{j=0}^{L-1}a_j\mathbf{C}_L^j =
\left[
\begin{matrix}
a_0 & a_1 & \ldots & a_{L-1} \\
a_{L-1} & a_0 & \ldots & a_{L-2} \\
\vdots & \ddots & \ddots & \vdots \\
a_1 & \ldots & a_{L-1} & a_0 \\
\end{matrix}
\right]
\end{equation}
it is qualified as a \emph{circulant matrix}. Thus, according to Lemma \ref{lemma:cyclic_matrix_decomposition} in \cite{Newman83}, for any $L \geq 1$, we have the following formula on the rank of $\sum_{j=0}^{L-1}a_j\mathbf{C}_L^j$:
\begin{equation}
\label{eqn:rank_circulant_matrix}
\mathrm{rank}\left(\sum\nolimits_{j=0}^{L-1}a_j\mathbf{C}_L^j\right) = L - \mathrm{deg}\left(g(x)\right),
\end{equation}
where $g(x)$ refers to the polynomial over GF($2$) that is the greatest common divisor of $x^L - 1$ and $\sum_{j=0}^{L-1}a_jx^j$, and $\mathrm{deg}(g(x))$ means the degree of $g(x)$.

\section{Algebraic Formulation of Circular-Shift LNC on a General Network}
Similar ideas of adopting circular-shifts and bit-wise additions as encoding operations have been respectively considered in \cite{Khojastepour10} and \cite{HouShum16} to model the rotation-and-add linear codes for a multicast network and the BASIC functional-repair regenerating codes for a distributed storage system. Their approach stems from the cyclic codes in coding theory, and relates the binary sequences transmitted on edges and the local encoding kernels to polynomials. %
Due to the lack of a systematic model, they showed the code existence but did not provide any algorithm for efficient code construction. %

We next model circular-shift LNC as a subclass of vector LNC, so that the local encoding kernels are particular circulant matrices prescribed by the set $\mathcal{C}_\delta$ in (\ref{eqn:set_C_delta}). The advantage of such formulation is that we can make use of Lemma \ref{lemma:cyclic_matrix_decomposition} to conduct more transparent manipulations on the matrix operations among local encoding kernels. %
An inherent connection between circular-shift LNC and scalar LNC can be subsequently established not only on a multicast network, but on a general network as well. As an application, it can facilitate efficient construction of circular-shift linear solutions for multicast networks.

\begin{definition}
\label{Def:Circular_Shift_LNC}
On a general network, an $(L', L)$ \emph{circular-shift linear code of degree} $\delta$ refers to an $(L', L)$-fractional linear code $(\mathbf{K}_{d,e})$ over GF($2$) with all local encoding kernels chosen $\mathbf{K}_{d,e}$ from $\mathcal{C}_\delta$ defined in (\ref{eqn:set_C_delta}). It is called an $(L', L)$ \emph{circular-shift linear solution of degree $\delta$} if it is an $(L', L)$-fractional linear solution. %
\end{definition}

It is interesting to note that the set $\mathcal{C}_L$ forms a \emph{commutative} subring of the (non-commutative) ring $M_L(\mathrm{GF}(2))$ of $L \times L$ binary matrices. Thus, circular-shift LNC conforms to the assumption in the algebraic structure of vector LNC that local encoding kernels are selected from commutative matrices \cite{Ebrahimi}. In addition, under the general model in \cite{Zeger17}, an $L$-dimensional (\emph{i.e.} $(L, L)$) circular-shift linear code of degree $L$ can be regarded as a linear code over the $\mathcal{C}_L$-module $\mathrm{GF}(2)^L$.

It is also worthwhile noting that rotation-and-add coding studied in \cite{Khojastepour10} can be regarded as a special type of circular-shift LNC of degree $1$, where matrix $\mathbf{0}$ is not a candidate for local encoding kernels.

Since every matrix in $\mathcal{C}_\delta$ is the summation of at most $\delta$ cyclic permutation matrices of size $L$, the operation $\mathbf{m}_d \mathbf{K}_{d,e}$ on an $L$-dimensional binary row vector $\mathbf{m}_d$ conducts at most $\delta$ circular-shifts and then computes bit-wise additions among at most $\delta$ circular-shifted row vectors.

\begin{figure}[!t]
\centering
\scalebox{0.73}
{\includegraphics{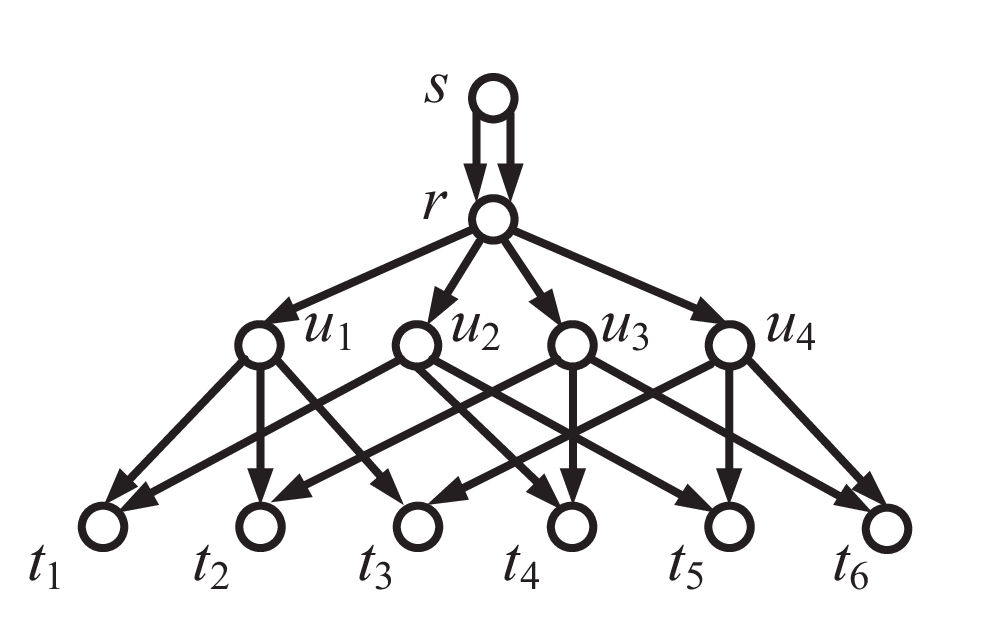}}
\caption{The $(4,2)$-Combination Network has a unique source with $\omega = 2$ and $6$ receivers at the bottom.}
\label{Fig:Combination_Network_4_2}
\end{figure}

\begin{example}
Fig. \ref{Fig:Combination_Network_4_2} depicts the $(4, 2)$-Combination Network, which is a multicast network with four layers. The top layer consists of the unique source $s$ with out-degree $2$, the third layer consists of $4$ nodes, and a bottom-layer receiver is connected from every pair of layer-3 nodes. Consider the following $(4, 5)$ circular-shift linear code $(\mathbf{K}_{d,e})$ of degree $1$. Denote by $\mathbf{m}_1' = [m_{11}~m_{12}~m_{13}~m_{14}]$ and $\mathbf{m}_2' = [m_{21}~m_{22}~m_{23}~m_{24}]$ the two data units generated at $s$. The data units transmitted on the two outgoing edges $e_1, e_2$ of $s$ are, respectively, $\mathbf{m}_1 = [0~m_{11}~m_{12}~m_{13}~m_{14}]$ and $\mathbf{m}_{2} = [0~m_{21}~m_{22}~m_{23}~m_{24}]$. The local encoding kernels for adjacent pairs $(e_i, ru_j)$ are
\begin{equation}
\label{eqn:example_4-2-Combination}
\begin{matrix}
\mathbf{K}_{e_1, ru_1} = \mathbf{K}_{e_1, ru_2} = \mathbf{K}_{e_1, ru_3} = \mathbf{K}_{e_1, ru_4} = \mathbf{K}_{e_2, ru_2} = \mathbf{I}_5, \\
\mathbf{K}_{e_2, ru_1} = \mathbf{0}, ~
\mathbf{K}_{e_2, ru_3} = \mathbf{C}_5,~
\mathbf{K}_{e_2, ru_4} = \mathbf{C}_5^2,
\end{matrix}
\end{equation}
and all local encoding kernels at nodes $u_j$, $1 \leq j \leq 4$, are the identity matrix $\mathbf{I}_5$. %
Thus, the binary sequence transmitted on every edge $ru_j$, $1 \leq j \leq 4$, can be computed as
\begin{align}
\begin{split}
\mathbf{m}_{ru_1} &=\mathbf{m}_1\mathbf{I}_{5} + \mathbf{m}_2\mathbf{0} = [0~m_{11}~m_{12}~ m_{13}~m_{14}], \\
\mathbf{m}_{ru_2} &= \mathbf{m}_1\mathbf{I}_{5} + \mathbf{m}_2\mathbf{I}_5 = [0~m_{11}+ m_{21}~m_{12}+ m_{22}~ m_{13}+ m_{23}~m_{14}+ m_{24}], \\
\mathbf{m}_{ru_3} &= \mathbf{m}_1\mathbf{I}_5+\mathbf{m}_2\mathbf{C}_5 = [m_{24}~ m_{11}~ m_{12}+ m_{21}~ m_{13}+ m_{22}~ m_{14}+ m_{23}], \\
\mathbf{m}_{ru_4} &= \mathbf{m}_1\mathbf{I}_5+\mathbf{m}_2\mathbf{C}_5^2 = [m_{23}~~ m_{11}+ m_{24}~~ m_{12}~~ m_{13}+ m_{21}~~ m_{14}+ m_{22}],
\end{split}
\end{align}
and $[\mathbf{m}_{e}]_{e\in \mathrm{In}(t_2)} = [\mathbf{m}_{ru_1}~\mathbf{m}_{ru_3}]$, $[\mathbf{m}_{e}]_{e\in \mathrm{In}(t_6)} = [\mathbf{m}_{ru_3}~\mathbf{m}_{ru_4}]$. For receiver $t_2$, given the $10\times 10$ binary matrix $\setlength{\arraycolsep}{2.0pt}
\renewcommand{\arraystretch}{0.7}\mathbf{D}_2 = \begin{bmatrix} \mathbf{I}_5 & \mathbf{C}_5^4 \\ \mathbf{0} & \mathbf{C}_5^4 \end{bmatrix}$, the circular-shift-based operations $[\mathbf{m}_{e}]_{e\in \mathrm{In}(t_2)}\mathbf{D}_{2}$ yields
\begin{align}
[\mathbf{m}_{e}]_{e\in \mathrm{In}(t_2)}\mathbf{D}_{2} &=
[\mathbf{m}_{ru_1}~~\mathbf{m}_{ru_1}\mathbf{C}_5^4+\mathbf{m}_{ru_3}\mathbf{C}_5^4]
= [\mathbf{m}_{1}~\mathbf{m}_{2}],
\end{align}
based on which the two source data units $\mathbf{m}_1'$, $\mathbf{m}_2'$ can be directly recovered. For receiver $t_6$, given the $10\times 10$ binary matrix $\setlength{\arraycolsep}{3.0pt}
\renewcommand{\arraystretch}{0.7}\mathbf{D}_6 =
\begin{bmatrix} \mathbf{C}^4_{5}+\mathbf{C}_{5}^2 & \mathbf{C}^2_{5}+\mathbf{I}_{5} \\ \mathbf{C}^3_{5}+\mathbf{C}_{5} & \mathbf{C}^2_{5}+\mathbf{I}_{5} \end{bmatrix}$, the circular-shift-based operations $[\mathbf{m}_{e}]_{e\in \mathrm{In}(t_6)}\mathbf{D}_{6}$ yields
\begin{align}
[\mathbf{m}_{e}]_{e\in \mathrm{In}(t_6)}\mathbf{D}_{6} &=
 \left[\mathbf{m}_{ru_3}(\mathbf{C}^4_{5}+\mathbf{C}_{5}^2)+\mathbf{m}_{ru_4}(\mathbf{C}^3_{5}+\mathbf{C}_{5}) ~~
 \mathbf{m}_{ru_3}(\mathbf{C}^2_{5}+\mathbf{I}_{5})+\mathbf{m}_{ru_4}(\mathbf{C}^2_{5}+\mathbf{I}_{5})
\right] \\
%&= \left[\mathbf{m}_{1}(\mathbf{C}^4_{5}+\mathbf{C}_{5}^2+\mathbf{C}^3_{5}+\mathbf{C}_{5})
%+\mathbf{m}_{2}(\mathbf{C}_{5}\mathbf{C}^4_{5}+\mathbf{C}_{5}\mathbf{C}_{5}^2+\mathbf{C}_{5}^2\mathbf{C}^3_{5}+\mathbf{C}_{5}^2\mathbf{C}_{5}) \right. \nonumber \\
%& ~~~~\left.\mathbf{m}_{1}(\mathbf{C}^2_{5}+\mathbf{I}_{5}+\mathbf{C}^2_{5}+\mathbf{I}_{5})+
%\mathbf{m}_2(\mathbf{C}_{5}\mathbf{C}^2_{5} + \mathbf{C}_{5} + \mathbf{C}^2_{5}\mathbf{C}^2_{5} + \mathbf{C}^2_{5})\right] \\
&=
[\mathbf{m}_{1}(\mathbf{C}_{5}+\mathbf{C}_{5}^2+\mathbf{C}^3_{5}+\mathbf{C}^4_{5})~~\mathbf{m}_{2}(\mathbf{C}_{5}+\mathbf{C}_{5}^2+\mathbf{C}^3_{5}+\mathbf{C}^4_{5})].
\end{align}
Note that $\mathbf{m}_{1}(\mathbf{C}_{5}+\mathbf{C}_{5}^2+\mathbf{C}^3_{5}+\mathbf{C}^4_{5}) = [m_1~m_1+m_{11}~m_1+m_{12}~m_1+m_{13}~m_1+m_{14}]$ and $\mathbf{m}_{2}(\mathbf{C}_{5}+\mathbf{C}_{5}^2+\mathbf{C}^3_{5}+\mathbf{C}^4_{5}) = [m_2~m_2+m_{21}~m_2+m_{22}~m_2+m_{23}~m_2+m_{24}]$, where $m_1 = \sum_{1\leq j \leq 4}m_{1j}$ and $m_2 = \sum_{1\leq j \leq 4}m_{2j}$. Thus, the two source data units $\mathbf{m}_1'$, $\mathbf{m}_2'$ can be conveniently recovered at $t_6$ from $[\mathbf{m}_{e}]_{e\in \mathrm{In}(t_6)}\mathbf{D}_{6}$ too. Analogously, one may check that for receivers $t_1, t_3, t_4, t_5$, the source data units can also be respectively recovered based on $\setlength{\arraycolsep}{2.0pt}
\renewcommand{\arraystretch}{0.7}
\mathbf{D}_1 =
\begin{bmatrix}
\mathbf{I}_{5} & \mathbf{I}_{5} \\
\mathbf{0} & \mathbf{I}_{5}
\end{bmatrix},
\mathbf{D}_3 =
\begin{bmatrix}
\mathbf{I}_{5} & \mathbf{C}^4_{5}+\mathbf{C}^2_{5} \\
\mathbf{0} & \mathbf{C}^4_{5}+\mathbf{C}^2_{5}
\end{bmatrix},
\mathbf{D}_4 =
\begin{bmatrix}
\mathbf{C}^4_{5}+\mathbf{C}^2_{5} & \mathbf{C}^3_{5}+\mathbf{C}_{5} \\
\mathbf{C}^3_{5}+\mathbf{C}_{5} & \mathbf{C}^3_{5}+\mathbf{C}_{5}
\end{bmatrix},
\mathrm{and}~
\mathbf{D}_5 =
\begin{bmatrix}
\mathbf{C}^4_{5}+\mathbf{C}^3_{5} & \mathbf{C}^2_{5}+\mathbf{C}_{5} \\
\mathbf{C}^2_{5}+\mathbf{C}_{5} & \mathbf{C}^2_{5}+\mathbf{C}_{5}
\end{bmatrix}.
$
In all, the considered code $(\mathbf{K}_{d,e})$ qualifies as a $(4, 5)$ circular-shift linear solution. \hfill $\blacksquare$
\end{example}

The $(4,5)$ circular-shift linear solution in the above example for the $(4,2)$-Combination Network is not coincidentally constructed. Let $\alpha \in \mathrm{GF}(2^4)$ be a root of the irreducible polynomial $f(x) = x^4+x^3+x^2+x+1$ over GF($2$). Since $f(x)$ divides $x^5 + 1$, $\alpha$ is a root of $x^5 + 1$ and thus $\alpha^5 = 1$. %
% so that the $16$ elements of GF($2^4$) can be represented as $\{a_3\alpha^3 + a_2\alpha^2 + a_1\alpha + a_0: a_0, a_1, a_2, a_3 \in \{0, 1\}\}$.
Via replacing $\mathbf{C}_5$ by $\alpha$ in (\ref{eqn:example_4-2-Combination}), we can obtain a counterpart scalar linear code $(k_{d,e})$ over GF($2^4$) prescribed by
\begin{equation}
\label{eqn:example_4-2-Combination-scalar}
k_{e_1, ru_1} = k_{e_1, ru_2} = k_{e_1, ru_3} = k_{e_1, ru_4} = k_{e_2, ru_2} = 1, k_{e_2, ru_1} = 0, k_{e_2, ru_3} = \alpha, k_{e_2, ru_4} = \alpha^2,
\end{equation}
and all local encoding kernels at nodes $u_j$, $1 \leq j \leq 4$, equal to $1$. %
For this scalar code, given that the two data units generated at $s$ are $m_1, m_2 \in \mathrm{GF}(2^4)$, the data units received by receiver $t_2$ and $t_6$ are $[m_e]_{e \in \mathrm{In}(t_2)} = [m_1~~m_1+\alpha m_2]$ and $[m_e]_{e \in \mathrm{In}(t_6)} = [m_1+\alpha m_2~~m_1+\alpha^2 m_2]$, respectively. Thus, $[m_e]_{e \in \mathrm{In}(t_2)}\mathbf{D}_2 = [m_e]_{e \in \mathrm{In}(t_6)}\mathbf{D}_6 = [m_1~m_2]$ with $\setlength{\arraycolsep}{2.0pt}
\renewcommand{\arraystretch}{0.6}\mathbf{D}_2 = \begin{bmatrix} 1 & \alpha^4 \\ 0 & \alpha^4 \end{bmatrix}$ and $\setlength{\arraycolsep}{2.0pt}
\renewcommand{\arraystretch}{0.6}\mathbf{D}_6 = \begin{bmatrix} \alpha^4+\alpha^2 & \alpha^2+1 \\ \alpha^3+\alpha & \alpha^2+1 \end{bmatrix}$. Similarly, one may further check that receiver $t_1, t_3, t_4, t_5$ can respectively recover $m_1, m_2$ from the received data units based on $\setlength{\arraycolsep}{2.0pt}
\renewcommand{\arraystretch}{0.6} \mathbf{D}_1 =
\begin{bmatrix}
1 & 1 \\
0 & 1
\end{bmatrix},
\mathbf{D}_3 =
\begin{bmatrix}
1 & \alpha^4 + \alpha^2 \\
0 & \alpha^4 + \alpha^2
\end{bmatrix},
\mathbf{D}_4 =
\begin{bmatrix}
\alpha^4 + \alpha^2 & \alpha^3 + \alpha \\
\alpha^3 + \alpha & \alpha^3 + \alpha
\end{bmatrix},
\mathrm{and}~
\mathbf{D}_5 =
\begin{bmatrix}
\alpha^4 + \alpha^3 & \alpha^2 + \alpha \\
\alpha^2 + \alpha & \alpha^2 + \alpha
\end{bmatrix}$. %
Hence, code $(k_{d,e})$ qualifies as a scalar linear solution.

We shall next show that the connection between the scalar linear solution over GF($2^4$) and the $(4, 5)$ circular-shift linear solution demonstrated above intrinsically holds between a scalar linear solution over GF($2^{L-1}$) and an $(L-1,L)$ circular-shift linear solution for an arbitrary network, given that $L$ is a prime with primitive root $2$, that is, the multiplicative order of $2$ modulo $L$ is equal to $L-1$. Such a condition on $L$ endows us with the following simple but useful propositions.

\begin{lemma}
\label{lemma:extension_field}
Let $L$ be a prime with primitive root $2$ and $\alpha$ be  a primitive $L^{th}$ root of unity over GF(2). The following hold:
\begin{enumerate}[a)]
  \item $f(x) = x^{L-1} + \ldots + x + 1$ is an irreducible polynomial over GF($2$) and it has $L-1$ roots: $\alpha, \ldots, \alpha^{L-1}$, which belong to GF($2^{L-1}$).
  \item Corresponding to every element $k \in \mathrm{GF}(2^{L-1})$, there is a unique polynomial over GF($2$)
    \begin{equation}
    \label{eqn:field-element-unique-expression-polynomial}
    g(x) := a_{L-1}x^{L-1} + \ldots + a_{1}x^1 + a_0,
    \end{equation}
    subject to $k = g(\alpha)$, and at most $\frac{L-1}{2}$ nonzero~coefficients $a_j$, $0 \leq j \leq L-1$.
%    \begin{equation}
%    \label{eqn:field-element-unique-expression}
%    k = %g(\alpha),~\mathrm{and~at~most}~\frac{L-1}{2}~\mathrm{nonzero~coefficients}~ a_j, 0 \leq j \leq L-1.
%    \end{equation}
  \item For two arbitrary polynomials $g_1(x)$ and $g_2(x)$ over GF($2$), if $g_1(\alpha^{k_1}) = g_2(\alpha^{k_2})$, then $g_1(\alpha^{jk_1}) = g_2(\alpha^{jk_2})$ for all $1 \leq j \leq L-1$.
\end{enumerate}
\begin{proof}
Though the application of (a) and (b) related to GF($2^{L-1}$) can also be found in \cite{HouShum16} and \cite{MDS_array_code}, we still provide the proof in Appendix-\ref{Appendix:Lemma_Extension_Field_Proof} for self sufficiency.
\end{proof}
\end{lemma}

\begin{notations}
Let $L$ be a prime with primitive root $2$, and $\alpha$ be a primitive $L^{th}$ root of unity over GF(2).

When an element in GF($2^{L-1}$) is expressed as $g(\alpha)$, $g(x)$ means a polynomial over GF($2$) in the form of (\ref{eqn:field-element-unique-expression-polynomial}) with at most $\frac{L-1}{2}$ nonzero terms. %
Similarly, when an $m \times n$ matrix over GF($2^{L-1}$) is expressed as $\mathbf{M}(\alpha)$, $\mathbf{M}(x)$ means a matrix over the polynomial ring GF(2)$[x]$, in which every entry is a polynomial in the form of (\ref{eqn:field-element-unique-expression-polynomial}) with at most $\frac{L-1}{2}$ nonzero terms. %
Further, $\mathbf{M}(\alpha^i)$, $i \geq 0$, represents the $m \times n$ matrix over GF($2^{L-1}$) obtained from $\mathbf{M}(x)$ via setting $x$ to $\alpha^i$, and $\mathbf{M}(\mathbf{C}_L^i)$ represents the $mL \times nL$ matrix over GF($2$) obtained from $\mathbf{M}(x)$ via replacing every zero entry by the $L \times L$ zero matrix and setting $x$ to be the matrix $\mathbf{C}_L^i$.
\end{notations}

On an arbitrary network, given a scalar linear code $(g_{d,e}(\alpha))$ over GF($2^{L-1}$), construct an $(L-1, L)$ circular-shift linear code $(\mathbf{K}_{d,e})$ as follows:
\begin{itemize}
\item for each $s \in S$, the data unit $\mathbf{m}_e$ transmitted on $e \in \mathrm{Out}(s)$ is $\mathbf{m}_e = [0~\mathbf{m}_e']$, where $\mathbf{m}_e'$ is one of the $|\mathrm{Out}(s)|$ $(L-1)$-dimensional binary row vectors generated at $s$.
\item for every adjacent pair $(d,e)$ of edges, the local encoding kernel $\mathbf{K}_{d,e}$ is
    \begin{equation}
    \label{eqn:circular-shift-LEK-Construction}
        \mathbf{K}_{d, e} = g_{d,e}(\mathbf{C}_L).
    \end{equation}
\end{itemize}
An inherent connection between the scalar linear code $(g_{d,e}(\alpha))$ and the circular-shift linear code $(\mathbf{K}_{d,e})$ is established by the following fundamental theorem of the present paper.

%Since there are at most $\frac{L-1}{2}$ nonzero terms in $g_{d,e}(x)$, $\mathbf{K}_{d, e} = g_{d,e}(\mathbf{C}_L) \in \mathcal{C}_{\frac{L-1}{2}}$, and so the constructed $(L-1, L)$-fractional linear code $(\mathbf{K}_{d,e})$ is of degree $\frac{L-1}{2}$.

\begin{theorem}
\label{thm:deterministic_construction}
If $(g_{d,e}(\alpha))$ is a scalar linear solution, then the constructed $(\mathbf{K}_{d,e})$ is an $(L-1, L)$ circular-shift linear solution of degree $\frac{L-1}{2}$, \emph{i.e.}, with all $\mathbf{K}_{d,e}$ belonging to $\mathcal{C}_{\frac{L-1}{2}}$ defined in (\ref{eqn:set_C_delta}). In addition, if $\mathbf{D}_t(\alpha)$ is the $|\mathrm{In}(t)| \times \omega_t$ decoding matrix for a receiver $t$, then the decoding matrix of $(\mathbf{K}_{d,e})$ for $t$ is given by
\begin{align}
\mathbf{D}_t(\mathbf{C}_L)
\cdot(\mathbf{I}_{\omega_t}\otimes\tilde{\mathbf{I}}_L),
\end{align}
where $\tilde{\mathbf{I}}_L$ denotes the $L \times (L-1)$ matrix obtained by inserting a row vector of all ones on top of $\mathbf{I}_{L-1}$.
\begin{proof}
The essence of the proof is to make use of the diagonalization manipulation on the local encoding kernels $\mathbf{K}_{d,e}$ based on Lemma \ref{lemma:cyclic_matrix_decomposition}, and the fact that the scalar linear code $(g_{d,e}(\alpha^j))$ is also a linear solution for all $1 \leq j \leq {L-1}$, which can be proved based on Lemma \ref{lemma:extension_field}. The details of the proof are given by Appendix-\ref{Appendix:Main-Thm-Proof}.
\end{proof}
\end{theorem}

One may observe that the mapping from $g_{d,e}(\alpha) \in \mathrm{GF}(2^{L-1})$ to $\mathbf{K}_{d,e} \in \mathcal{C}_{\frac{L-1}{2}}$ used in (\ref{eqn:circular-shift-LEK-Construction}) for code construction is a one-to-one correspondence.
However, such a mapping is \emph{not} an isomorphism because $\mathcal{C}_{\frac{L-1}{2}}$ is not closed under matrix addition, and some matrix in $\mathcal{C}_{\frac{L-1}{2}}$ (e.g., $\mathbf{I}_L + \mathbf{C}_L$) is not invertible. This makes the established intrinsic connection between circular-shift LNC and scalar LNC non-trivial.

It turns out that when $L$ is a prime with primitive root $2$, as long as a general network has a scalar linear solution over GF($2^{L-1}$), it has an alternative $(L-1, L)$ circular-shift linear solution of degree $(L-1)/2$ too. %
Different from previous studies in \cite{Jaggi06}-\cite{Khojastepour10}, which mainly consider low complexity encoding operations, the constructed $(L-1,L)$ circular-shift linear solution builds up not only local encoding kernels, but also the decoding matrix based on cyclic permutation matrices.
% Actually, if a general network has an $(L-1, L)$ circular-shift linear solution, it has an scalar linear solution over GF($2^{L-1}$) too, as established in the next theorem.

\section{Deterministic Circular-Shift LNC on Multicast Networks}
\subsection{Deterministic Construction}
In the previous section, we have proved that for a general network, every scalar linear solution over {GF($2^{L-1}$), where $L$ is a prime with primitive root 2, can induce an $(L - 1, L)$ circular-shift linear solution of degree $\frac{L-1}{2}$. In this section, we restrict our attention to further investigate circular-shift LNC on multicast networks. Herein, unless otherwise specified, we still assume that $L$ is a prime with primitive root $2$. Unlike a general network, which may not have a linear solution over any module alphabet \cite{Zeger16}, there are various known algorithms, such as the ones in \cite{Jaggi05} and \cite{Langberg09}, to efficiently construct a scalar linear solution for a multicast network. Thus, as revealed by the next corollary, for a long enough block length $L$, an $(L - 1, L)$ circular-shift linear solution of an arbitrary degree can be efficiently constructed for every multicast network.

\begin{corollary}
\label{Cor:Efficient_Construction}
Let $1 \leq \delta \leq \frac{L-1}{2}$. For a multicast network, an $(L-1, L)$ circular-shift linear solution of degree $\delta$ can be efficiently constructed if %
the prime $L$ with primitive $2$ satisfies %
${\small
\left(\renewcommand{\arraystretch}{0.6}\begin{matrix} L \\ 0 \end{matrix} \right) +
\renewcommand{\arraystretch}{0.6}\left(\begin{matrix} L \\ 1\end{matrix} \right) + \ldots +
\renewcommand{\arraystretch}{0.6}\left(\begin{matrix} L \\ \delta \end{matrix} \right)} \geq |T|
$. %
\begin{proof}
By Lemma \ref{lemma:extension_field}.a), GF($2^{L-1}$) contains a primitive $L^{th}$ root of unity, which will be denoted by $\alpha$. Let $\mathcal{C}$ be a set of elements in GF($2^{L-1}$) which can be expressed in the form $a_0 + a_1\alpha + \ldots + a_{L-1}\alpha^{L-1}$ such that at most $\delta$ nonzero binary coefficients $a_j$, $0 \leq j \leq L-1$, are nonzero. Lemma \ref{lemma:extension_field}.b) implies that $\mathcal{C}$ contains ${\small \renewcommand{\arraystretch}{0.6}\left(\begin{matrix} L \\ 0 \end{matrix} \right) + \left(\begin{matrix} L \\ 1 \end{matrix} \right) + \ldots + \left(\begin{matrix} L \\ \delta \end{matrix} \right)}$ distinct elements. Then, if $|\mathcal{C}|$ is no smaller than the number of receivers, a scalar linear solution over $\mathrm{GF}(2^{L-1})$ can be efficiently constructed by the algorithm in \cite{Jaggi05} with local encoding kernels selected from $\mathcal{C}$. Thus, by Theorem \ref{thm:deterministic_construction}, it directly induces an $(L-1, L)$ circular-shift linear solution as well as the concomitant decoding matrix at every receiver.
\end{proof}
\end{corollary}

It is interesting to note that when the prime $L$ with primitive $2$ is larger than the number $|T|$ of receivers, the work in \cite{Khojastepour10} has proved that there \emph{exists} an $(L-1, L)$ circular-shift linear solution of degree $1$ for a multicast network. In addition, as the construction of a functional-repair regenerating code for a distributed storage system is essentially same as the construction of a scalar linear solution for a special multicast network (See, e.g., \cite{Dimakis-TIT}), the work (Theorem 7) in \cite{HouShum16} essentially proved the \emph{existence} of an $(L-1, L)$ circular-shift linear solution of degree $\frac{L-1}{2}$ for certain multicast networks. However, how to efficiently construct such desired circular-shift linear solutions was not known. Corollary \ref{Cor:Efficient_Construction} unveiled that all such desired circular-shift linear solutions can be efficiently constructed.

It is well-known (See, e.g., \cite{Zeger16}) that LNC over an arbitrary module alphabet is not sufficient to achieve the exact capacity of some (multi-source multicast) networks. As circular-shift LNC is a special class of vector LNC, it is not sufficient to achieve the exact capacity of these networks either. In contrast, for every multicast network, both scalar and vector LNC, over a long enough block length, can achieve the exact network capacity. Naturally, one may ask whether circular-shift LNC can achieve the exact capacity of every multicast network too. We next give a negative answer to it by demonstrating two instances.

Fig. \ref{Fig:Combination_Network} and Fig. \ref{Fig:Swirl_Network} respectively depict the classical $(n, 2)$-Combination Network (See, e.g., \cite{Ngai_Yeung}\cite{Xiao07}) and the Swirl Network recently designed in \cite{Sun_TIT15}. %
As a generalization of the $(4, 2)$-Combination Network depicted in Fig. \ref{Fig:Combination_Network_4_2}, there are also four layers of nodes in the $(n, 2)$-Combination Network, where the first layer consists of the unique source $s$, the third layer consists of $n$ nodes, and a bottom-layer receiver is connected from every pair of layer-3 nodes. %
It is known (See, e.g., \cite{Sun_TCom16}\cite{Etzion16}) that the $(n, 2)$-Combination Network has an $L$-dimensional vector linear solution over GF($2$) if and only if $2^L \geq n-1$. %
In addition, when $L$ is a prime no smaller than $n$, the work in \cite{Xiao07} proposed an interesting low-complexity $L$-dimensional LNC scheme for the $(n, 2)$-Combination Network based on circular-shifts together with a bit truncation process. Essentially, this scheme can be regarded as an $(L-1)$-dimensional vector linear solution over GF($2$) with all nonzero local encoding kernels equal to some $\hat{\mathbf{I}}\mathbf{C}_L^j\hat{\mathbf{I}}^T$, $0 \leq j \leq L-1$, where $\hat{\mathbf{I}}$ represents the $(L-1) \times L$ matrix $[\mathbf{I}_{L-1}~\mathbf{0}]$ obtained by appending a zero column vector after $\mathbf{I}_{L-1}$, and $\mathbf{C}_L$ is the cyclic permutation matrix defined in (\ref{eqn:cyclic_permutation_matrix}). %
The Swirl Network with the parameter $\omega \geq 3$ consists of five layers of nodes, where the top layer consists of the source node, each of the second and third layer consists of $\omega$ nodes, there are two layer-4 nodes connected from every layer-3 node, and a bottom-layer receiver is connected from every set $N$ of $\omega$ layer-4 nodes with the maximum flow from the source to $N$ equal to $\omega$. %
According to \cite{Sun_TCom16}, for every block length $L \geq 8$, the Swirl Network has an $L$-dimensional vector linear solution over GF(2). In contrast, the next proposition shows that if the local encoding kernels are restricted to be chosen from the set $\mathcal{C}_L$ of circulant matrices defined in (\ref{eqn:set_C_delta}), neither the $(n, 2)$-Combination Network nor the Swirl Network has an $L$-dimensional vector linear solution over GF(2) for any $L$.

\begin{figure}[!t]
\centering
\scalebox{0.63}
{\includegraphics{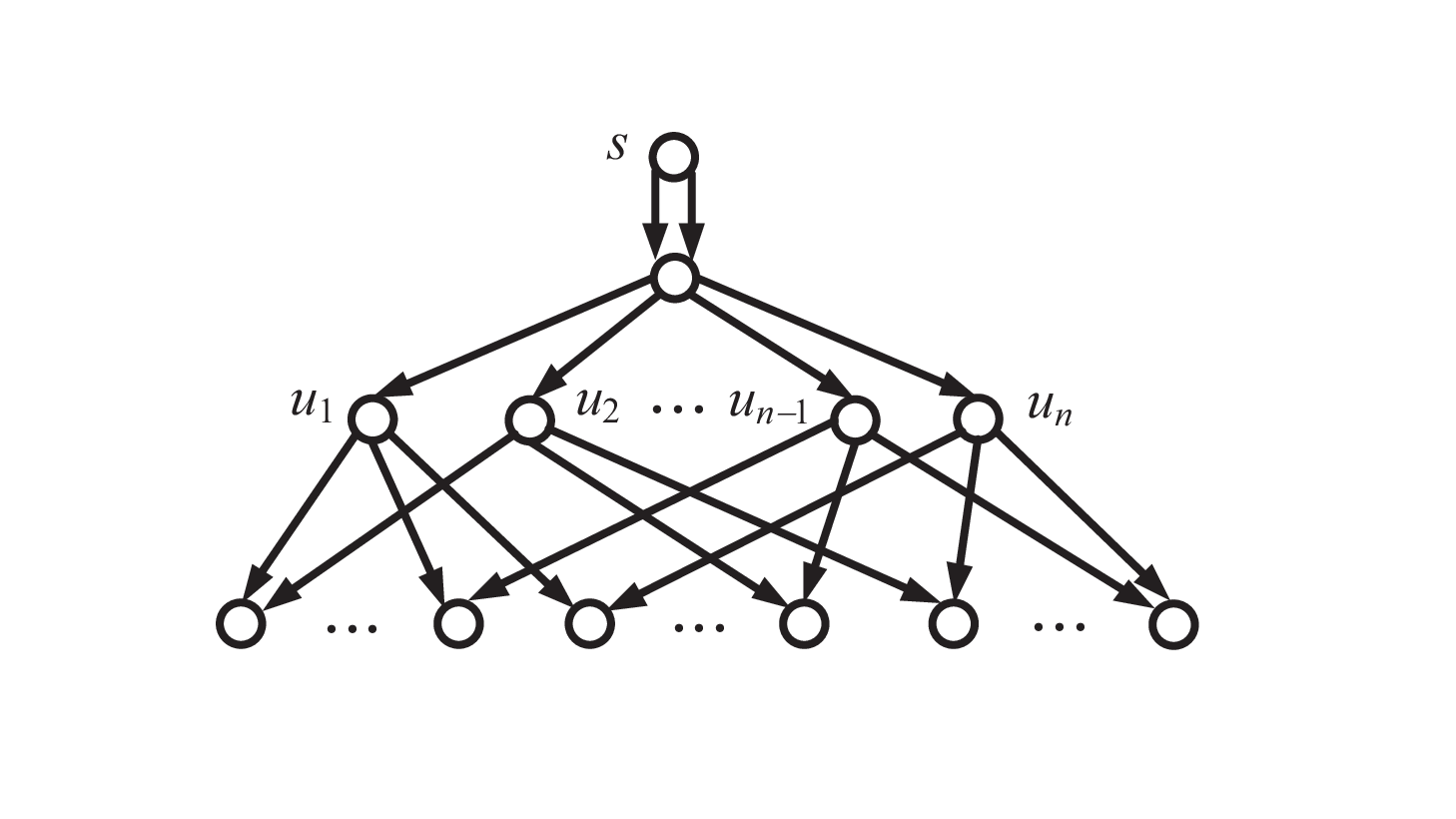}}
\caption{The classical $(n,2)$-Combination Network is known to have an $L$-dimensional vector linear solution over GF($2$) if and only if $2^L \geq n-1$.}
\label{Fig:Combination_Network}
\end{figure}

\begin{figure}[!t]
\centering
\scalebox{0.53}
{\includegraphics{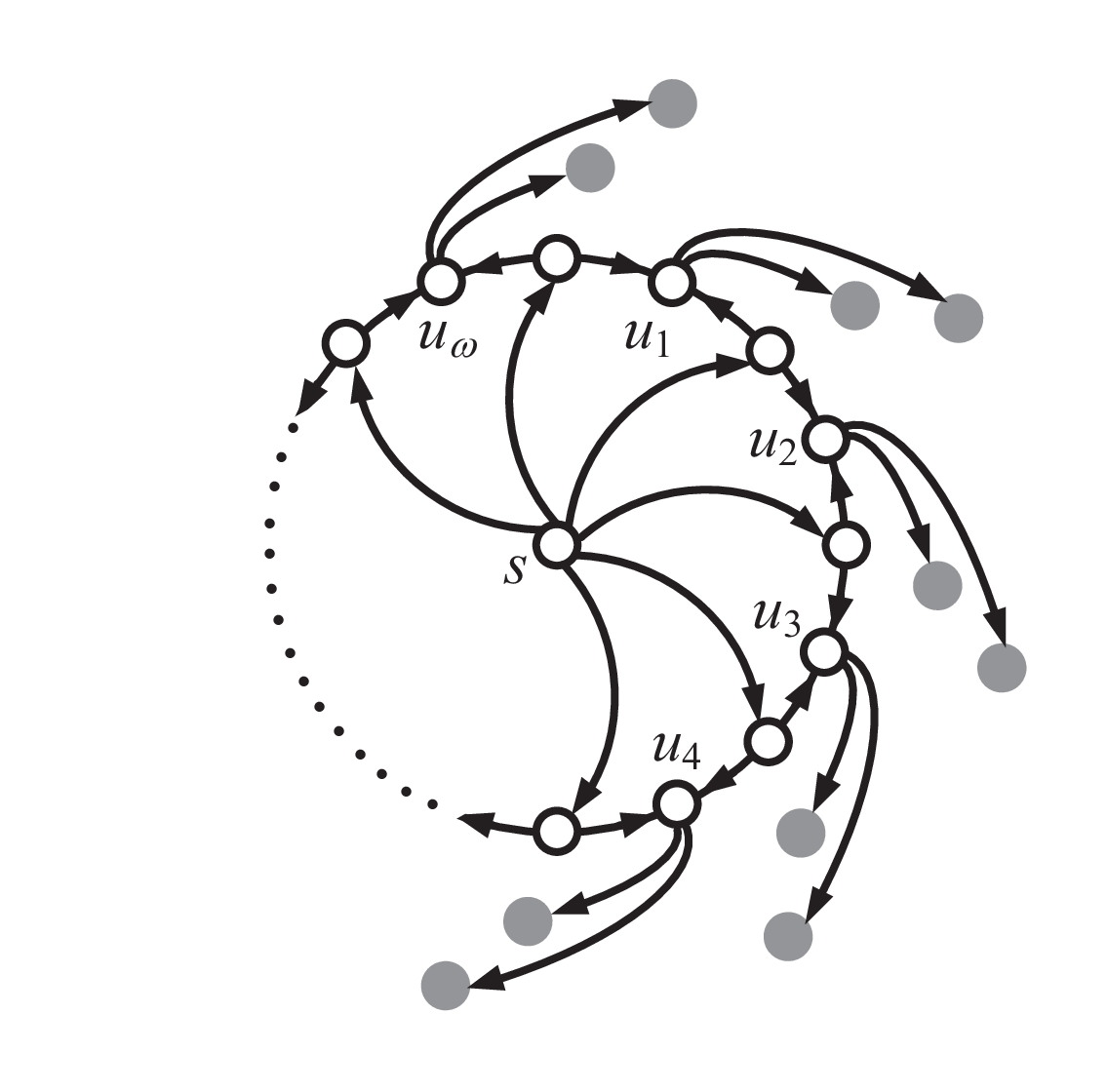}}
\caption{The Swirl Network with parameter $\omega$ has a non-depicted receiver connected from every set $N$ of $\omega$ grey nodes with the maximum flow from $s$ to $N$ equal to $\omega$. It has an $L$-dimensional vector linear solution over GF($2$) when $L \geq 8$.}
\label{Fig:Swirl_Network}
\end{figure}

\begin{proposition}
\label{prop:Insufficiency_Circular_Shift}

For $n\geq 4$, neither the $(n, 2)$-Combination Network nor the Swirl Network with parameter $\omega = n$ is $(L, L)$ circular-shift linearly solvable of degree $L$ for \emph{any} $L \geq 1$.
\begin{proof}
As implied from Equation $(4)$ and $(6)$ in \cite{Sun_TCom16}, when $n \geq 4$, a necessary condition for both the $(n, 2)$-Combination Network and the Swirl Network with parameter $\omega = n$ to have an $L$-dimensional vector linear solution over GF($2$) is that there are two $L \times L$ invertible matrices $\mathbf{A}_1, \mathbf{A}_2$ over GF($2$) such that
\begin{equation}
\label{eqn:fullrank_circulant_matrix_sum}
\mathrm{rank}(\mathbf{A}_i + \mathbf{A}_j) = L.
\end{equation} %
Let $\mathbf{A}_1 = \sum_{j=0}^{L-1}a_{1j}\mathbf{C}_L^j$, $\mathbf{A}_2 = \sum_{j=0}^{L-1}a_{2j}\mathbf{C}_L^j$ be two invertible matrices in $\mathcal{C}_L$. %
According to (\ref{eqn:rank_circulant_matrix}) in Section II.B, $\mathrm{rank}(\mathbf{A}_1) = L - \mathrm{deg}(g_1(x))$, where $g_1(x)$ is the greatest common divisor of $x^L - 1$ and $\sum_{j=0}^{L-1} a_{1j}x^j$. If there are an even number of nonzero coefficients among $a_{1j}$, $0 \leq j \leq L-1$, then $x^L - 1$ and $\sum_{j=0}^{L-1} a_{1j}x^j$ have a common root $1$, so $x-1$ divides $g_1(x)$ and $\mathrm{rank}(\mathbf{A}_1) < L$, a contradiction to that $\mathbf{A}_1$ is invertible. Therefore, there are an odd number of nonzero coefficients among $a_{1j}$, $0 \leq j \leq L-1$. Similarly, there are an odd number of nonzero coefficients among $a_{2j}$, $0 \leq j \leq L-1$, too.
As a result, the number of nonzero coefficients among $a_{1j}+a_{2j}~(\in \mathrm{GF}(2)), 0 \leq j \leq L-1$ must be even. This in turn implies that $x - 1$ divides both $x^L - 1$ and $\sum_{j=0}^{L-1} (a_{1j}+a_{2j})x^j$, so $\mathbf{A}_1 + \mathbf{A}_2 = \sum_{j=0}^{L-1}(a_{1j} + a_{2j})\mathbf{C}_L^j$ cannot be full rank. We can then conclude that neither the $(n,2)$-Combination Network nor the Swirl Network is $(L, L)$ circular-shift linearly solvable of degree $L$ for any $L \geq 1$.
\end{proof}
\end{proposition}

Proposition \ref{prop:Insufficiency_Circular_Shift} justified the optimality of the $(L-1, L)$ circular-shift linear solution efficiently constructed in Corollary \ref{Cor:Efficient_Construction} for an arbitrary multicast network, in the sense that the 1-bit redundancy is inevitable.

According to Artin's conjecture on primitive roots (See, e.g., \cite{Sloane_Prime_List}), there are infinitely many primes with primitive root $2$. While the conjecture is open, there are sufficiently many such primes $L$ (See the table in \cite{Sloane_Prime_List}) to choose for efficient construction of an $(L-1, L)$ circular-shift linear solution for a multicast network. %

\subsection{Computational Complexity Comparison}
We now compare the encoding and decoding complexity between circular-shift LNC and scalar LNC, from the perspective of required binary operations. %
To keep the same benchmark for complexity comparison, we adopt the following assumptions similar to in \cite{HouShum16}. We shall ignore the complexity of a circular-shift operation on a binary sequence, which can be software implemented by modifying the pointer to the starting address in the sequence, and we just consider the standard implementation of multiplication in GF($2^L$) by polynomial multiplication modulo an irreducible polynomial, instead of considering other advanced techniques such as the FFT algorithm  \cite{fast_FFT}.

On a multicast network, let $v$ be an intermediate node with indegree $\eta$, and $t \in T$ be a receiver. First consider a scalar linear solution over GF($2^L$). Node $v$ takes $\eta$ multiplications and $\eta-1$ additions over GF($2^L$) to generate the data symbol $m_e = \sum_{d\in\mathrm{In}(v)} m_dk_{d,e} \in \mathrm{GF}(2^L)$ for an outgoing edge $e \in \mathrm{Out}(v)$. Receiver $t$ takes $\omega^2$ multiplications and $\omega(\omega-1)$ additions over GF($2^L$) in the decoding process to recover $\omega$ source data symbols. When two elements in GF($2^L$) are expressed as two polynomials $f_1(x), f_2(x)$ of degree $L-1$ over GF($2$), it takes $L^2$ binary multiplications and $L(L-1)$ binary additions to compute $f_1(x)f_2(x)$. It takes additional $(L-1)(\kappa-1)$ binary operations to obtain $f_1(x)f_2(x)$ modulo $g(x)$, where $\kappa \geq 3$ represents the number of nonzero coefficients in $g(x)$. In total, node $v$ takes at least $\eta(2L^2+L)$ binary operations to obtain the $L$-bit data symbol $m_e$, and receiver $t$ takes at least $\omega^2 L(2L+1)$ binary operations to recover $\omega$ $L$-bit source data symbols.

Next consider an $L$-dimensional vector linear solution over GF($2$). In order to generate the data unit $\mathbf{m}_e = \sum_{d\in\mathrm{In}(v)} \mathbf{m}_d\mathbf{K_{d,e}} \in \mathrm{GF}(2)^L$ for an outgoing edge $e \in \mathrm{Out}(v)$, node $v$ takes $\eta L^2$ binary multiplications and $\eta L(L-1)+ (\eta-1)L = \eta L^2 - L$ binary additions. Receiver $t$ takes $\omega^2L^2$ binary multiplications and $\omega^2 L^2 - \omega L$ binary additions in the matrix operation $[\mathbf{m}_e]_{e\in \mathrm{In}(t)}\mathbf{D}_t$ to recover the $\omega L$ source data units.

Last consider an $(L-1, L)$ circular-shift linear solution of degree $\delta$ constructed by Theorem \ref{thm:deterministic_construction}. Node $v$ takes $L\left(\delta\eta - 1\right)$ binary operations to obtain the $L$-dimensional binary row vector $\mathbf{m}_e = \sum_{d \in \mathrm{In}(v)} \mathbf{m}_d\mathbf{K}_{d,e}$ for $e \in \mathrm{Out}(v)$. For receiver $t$, recall that the decoding matrix in Theorem \ref{thm:deterministic_construction} is given by
$\mathbf{D}_t(\mathbf{C}_L)\cdot(\mathbf{I}_\omega\otimes\tilde{\mathbf{I}}_L)$, where every block entry in the matrix $\mathbf{D}_t(\mathbf{C}_L)$ can be written as $\sum_{j=0}^{L-1}a_j\mathbf{C}_L^j$ with at most $\frac{L-1}{2}$ nonzero coefficients $a_j$. Thus, it takes $L\left(\frac{L-1}{2}\omega - 1\right)\omega$ binary operations to compute $[\mathbf{m}_e]_{e\in \mathrm{In}(t)}\cdot\mathbf{D}_t(\mathbf{C}_L)$ and additional $\omega L$ binary operations to further obtain $[\mathbf{m}_e]_{e\in \mathrm{In}(t)}\cdot\mathbf{D}_t(\mathbf{C}_L)\cdot(\mathbf{I}_\omega\otimes\tilde{\mathbf{I}}_L)$. In total, the number of binary operations is $\frac{1}{2}\omega^2L(L-1)$.

To the best of our knowledge, by known efficient algorithms in the literature, for an arbitrary multicast network, the minimum block length $L$, as a function of $|T|$, to respectively construct a scalar linear solution over GF($2^L$) and an $L$-dimensional vector linear solution over GF($2$) is the same $\lceil \log_2(|T|) \rceil$. In addition, according to Corollary \ref{Cor:Efficient_Construction}, the minimum block length $L$, which needs be a prime with primitive root $2$, to efficiently construct an $(L-1, L)$ circular-shift linear solution of degree $1$ and an $(L-1, L)$ circular-shift linear solution of degree $\frac{L-1}{2}$ is $|T|-1$ and $\lceil \log_2(|T|) \rceil + 1$, respectively. %
Therefore, to make a more transparent and fairer comparison, consider a scalar linear solution over GF($2^{m}$), %

an $m$-dimensional vector linear solution over GF(2), %
an $(m, m+1)$ circular-shift linear solution of degree $\frac{m}{2}$, and an $(L, L+1)$ circular-shift linear solution of degree $1$, where $m+1$, $L+1$ are primes with primitive root 2 and $2^{m} \geq L+2 \geq |T|$. %
 In this setting, all these four linear solutions can be efficiently constructed by known algorithms for an arbitrary multicast network. %
Table I lists the respective number of binary operations per bit for encoding at $v$ and decoding at $t$.
so that all these four linear solutions can be efficiently constructed by known algorithms for an arbitrary multicast network.

\begin{table}[!htbp]
\caption{Number of binary operations per bit for encoding and decoding}
\label{Table_Binary_Operation_Comparison}
\begin{center}
\begin{tabular}{|c|c|c|}

  \hline
  & Encoding & Decoding \\
  \hline
  Scalar over $\mathrm{GF}(2^m)$    & $> 2\eta m$                               &         $> \omega(2m+1)$                                              \\
  \hline
  $m$-dimensional vector  &  \centering{\multirow{2}*{$2\eta m -1$}}    &         \multirow{2}*{$2\omega m-1$}                  \\
  over GF(2)  & & \\
  \hline
  $(m, m+1)$ circular-shift        &  \centering{\multirow{2}*{$\frac{1}{2}\eta m$}}    &         \multirow{2}*{$\frac{1}{2}\omega (m+1)$}                  \\
  of degree $\frac{m}{2}$ & & \\
  \hline
  $(L, L+1)$ circular-shift   & \multirow{2}*{$\eta - 1$}             & \multirow{2}*{$\frac{1}{2}\omega(L+1) < \frac{1}{2}\omega 2^m$}    \\
  of degree $1$ & &  \\
  \hline
\end{tabular}
\end{center}
\end{table}

It can be seen that for the considered circular-shift linear solution of degree $\delta = \frac{m}{2}$, the number of required binary operations per bit for both encoding and decoding can be approximately reduced by $3/4$ compared with the scalar linear solution. When the degree of the circular-shift linear solution decreases from $\frac{m}{2}$ to $1$, the encoding complexity will decrease and the decoding complexity will increase. To our knowledge, this interesting tradeoff between encoding and decoding complexities for efficient construction of LNC schemes are new, and it makes circular-shift LNC more flexible to be applied in networks with different computational constraints.

One may observe that for the two circular-shift linear solutions in Table I, when $\delta$ decreases from $\frac{m}{2}$ to $1$, the increasing rate of the decoding complexity is faster than the decreasing rate of the encoding complexity. The reason is that for the method proposed in this paper, the necessary block length $m+1$ for efficiently constructing a circular-shift linear solution of degree $\frac{m}{2}$ is $\lceil\log_2{|T|}\rceil$, but the necessary block length $L+1$ for efficiently constructing a circular-shift linear solution of degree $1$ is $|T|$. How to efficiently construct a circular-shift linear solution of degree $1$ with a shorter block length deserves further investigation in future work.

\section{Random Circular-shift LNC on Multicast Networks}
\subsection{Probabilistic Analysis}
As we have not known whether there are infinitely many primes with primitive root $2$ yet, the results established in Theorem \ref{thm:deterministic_construction} and Corollary \ref{Cor:Efficient_Construction} are insufficient to imply that every multicast network is \emph{asymptotically circular-shift linearly solvable}, that is, for any $\epsilon > 0$, it has an $(L', L)$ circular-shift linear solution with $L'/L > 1 - \epsilon$. This motivates us to further study circular-shift LNC by random coding and to show, from a probabilistic perspective, that every multicast network is asymptotically circular-shift linearly solvable. With this aim, it suffices to consider circular-shift LNC of degree $1$, that is, all local encoding kernels are chosen from $\mathcal{C}_1 = \{\mathbf{0}, \mathbf{I}_L, \mathbf{C}_L, \ldots, \mathbf{C}^{L - 1}_L\}$. %
We first introduce the following lemma that will be useful in the analysis of the asymptotic linear solvability of random circular-shift LNC.

\begin{lemma}
\label{lemma:L+K}
For an $L \times L$ matrix $\mathbf{K}$ uniformly and randomly chosen from $\{\mathbf{0},\mathbf{I}_L, \mathbf{C}_L, \ldots, \mathbf{C}^{L - 1}_L\}$, an arbitrary $L\times L$ binary matrix $\mathbf{L}$, and an arbitrary real number $\epsilon>0$, the probability for the rank of $\mathbf{L}+\mathbf{K}$ lower than $L(1-\epsilon)$ is upper bounded by
\begin{equation}
\mathrm{Pr}\left(\mathrm{rank}(\mathbf{L}+\mathbf{K}) < L(1-\epsilon)\right) < 2^{-L\epsilon+\mathrm{log}(L+1)}.
\end{equation}
\begin{proof}
See Appendix-\ref{Appendix:Lemma_Proof}.
\end{proof}
\end{lemma}

We next consider the following way to randomly construct an $(L', L)$ circular-shift linear code:
\begin{itemize}
\item The $\omega L' \times \omega L$ coding matrix $\mathbf{G}_s$ operated at source $s$ is uniformly and randomly chosen from all $\omega L' \times \omega L$ binary matrices.
\item Every local encoding kernel is uniformly and randomly chosen from $\mathcal{C}_1 = \{\mathbf{0}, \mathbf{I}_L, \mathbf{C}_L, \ldots, \mathbf{C}^{L-1}_L\}$.
\end{itemize}

\begin{theorem}
\label{thm:asymptotic_solvable}
For every positive integer $L$, let $\epsilon_L > 0$ be an associated real number such that $\lim_{L\rightarrow\infty}\epsilon_L=0$ and $\lim_{L\rightarrow\infty} \log \cfrac{2^{L \epsilon_L}}{L+1} =\infty$, and let $L' = \cfrac{\omega - |E|\epsilon_L}{\omega}L$. The probability of a randomly constructed $(L', L)$ circular-shift linear code to be an $(L', L)$ linear solution is greater than $1-2^{-L\epsilon_L+\log(L+1)+\log|T||E|}$.
\begin{proof}
First, observe that for every receiver $t$, if $\mathrm{rank}(\mathbf{G}_s [\mathbf{F}_e]_{e \in \mathrm{In}(t)}) \geq \omega L'$, then there must exist an $\omega L \times \omega L'$ matrix $\mathbf{D}_t$ over GF(2) such that $\mathbf{G}_s [\mathbf{F}_e]_{e \in \mathrm{In}(t)} \mathbf{D}_t = \mathbf{I}_{\omega L'}$, that is, receiver $t$ can successfully recover the $\omega L'$ source data symbols. Thus, the probability of the randomly constructed code to be an $(L', L)$-fractional linear solution is lower bounded by
\begin{align}
\nonumber
& \mathrm{Pr}(\mathrm{rank}(\mathbf{G}_s [\mathbf{F}_e]_{e \in \mathrm{In}(t)}) \geq \omega L') \\
\geq & \mathrm{Pr}(\mathrm{rank}([\mathbf{F}_e]_{e \in \mathrm{In}(t)}) \geq r) \cdot \mathrm{Pr}(\mathrm{rank}(\mathbf{G}_s [\mathbf{F}_e]_{e \in \mathrm{In}(t)}) \geq \omega L'| \mathrm{rank}([\mathbf{F}_e]_{e \in \mathrm{In}(t)}) \geq r)
\end{align}
for an arbitrary $r \geq \omega L'$.

Consider an arbitrary receiver $t$ in the multicast network. As the maximum flow for $t$ is $\omega$, there are $\omega$ edge-disjoint paths from $s$ to $t$. Let $E_t \subset E$ denote the set of edges in the $\omega$ edge-disjoint paths and index the edges in $E_t$ as $e_1, e_2, \ldots, e_{|E_t|}$. Assume that there is an upstream-to-downstream order of $E_t$ with $\{e_1, \ldots, e_\omega\} = \mathrm{Out}(s)$ and $\{e_{|E_t|-\omega+1}, \ldots, e_{|E_t|}\} = \mathrm{In}(t)$. %
Iteratively consider an set $E_\omega$, which always consists of $\omega$ consecutive edges in $E_t$. Initially, $E_\omega = \{e_1, \ldots, e_\omega\}$ and by definition, $[\mathbf{F}_e]_{e\in E_\omega} = \mathbf{I}_{\omega L}$. In $i^{th} \geq 1$ iteration, based on the current setting $E_\omega$ which contains $e_{i+\omega -1}$ as the least ordered edge, define a new set $E_\omega' = E_\omega \backslash \{e_j\} \cup \{e_{i+\omega}\}$, where $(e_j, e_{i+\omega})$ forms an adjacent pairs of edges. Based on Lemma \ref{lemma:L+K}, it can be deduced (See Appendix-\ref{Appendix:Bound_Justification} for the details) that
%\begin{align}
%\label{eqn:Thm_Proof_Bound_0}
%& Pr(rank([\mathbf{F}_e]_{e\in E_\omega}) - rank([\mathbf{F}_e]_{e\in E_\omega'}) > L\epsilon_L) \nonumber \\
%\leq & 2^{-L \epsilon_L+\log(L+1)}.
%\end{align}
\begin{equation}
\label{eqn:Thm_Proof_Bound_0}
\mathrm{Pr}(\mathrm{rank}([\mathbf{F}_e]_{e\in E_\omega}) - \mathrm{rank}([\mathbf{F}_e]_{e\in E_\omega'}) > L\epsilon_L) \leq 2^{-L \epsilon_L+\log(L+1)}.
\end{equation}
Then, reset $E_\omega$ equal to $E_\omega'$ and proceed to the next iteration. In the final iteration, $E_\omega = \mathrm{In}(t)$. As the number of iterations conducted for $E_\omega$ to change from $\mathrm{Out}(s)$ to $\mathrm{In}(t)$ is upper bounded by $|E| - \omega$, the following can be readily obtained by a union bound on (\ref{eqn:Thm_Proof_Bound_0}):
%\begin{align}
%& Pr\{rank([\mathbf{F}_e]_{e \in In(t)}) \geq r\} \nonumber \\
%\geq &(1 - 2^{-L \epsilon_L+\log(L+1)})^{|E| - \omega} \nonumber \\
%\label{eqn:Thm_Proof_Bound_1}
%> &1 - (|E| - \omega) \cdot 2^{-L \epsilon_L+\log(L+1)},
%\end{align}
\begin{equation}
\label{eqn:Thm_Proof_Bound_1}
\mathrm{Pr}(\mathrm{rank}([\mathbf{F}_e]_{e \in \mathrm{In}(t)}) \geq r)
\geq (1 - 2^{-L \epsilon_L+\log(L+1)})^{|E| - \omega}
> 1 - (|E| - \omega) \cdot 2^{-L \epsilon_L+\log(L+1)},
\end{equation}
where $r$ is set to be $\omega L - L\epsilon_L (|E| - \omega)$.

Under the condition that $\mathrm{rank}([\mathbf{F}_e]_{e \in \mathrm{In}(t)}) \geq r$, it can be further deduced (See Appendix-\ref{Appendix:Bound_Justification} for the details) that
%\begin{align}
%\label{eqn:Thm_Proof_Bound_2}
%&Pr\{rank(\mathbf{G}_s [\mathbf{F}_e]_{e \in In(t)}) \geq \omega L' ~|~rank([\mathbf{F}_e]_{e \in In(t)}) \geq r\} \nonumber \\
%> & 1 - \omega L' 2^{-\omega L \epsilon_L}.
%\end{align}
\begin{equation}
\label{eqn:Thm_Proof_Bound_2}
\mathrm{Pr}(\mathrm{rank}(\mathbf{G}_s [\mathbf{F}_e]_{e \in \mathrm{In}(t)}) \geq \omega L' ~|~\mathrm{rank}([\mathbf{F}_e]_{e \in \mathrm{In}(t)}) \geq r) \\
> 1 - \omega L' 2^{-\omega L \epsilon_L}.
\end{equation}
Then, by combining (\ref{eqn:Thm_Proof_Bound_1}) and (\ref{eqn:Thm_Proof_Bound_2}),
%\begin{align}
%& \mathrm{Pr}(\mathrm{rank}(\mathbf{G}_s [\mathbf{F}_e]_{e \in \mathrm{In}(t)}\geq\omega L') \nonumber \\
%> & 1 - (|E| - \omega) \cdot 2^{-L \epsilon_L+\log(L+1)} - \omega L' 2^{-\omega L \epsilon_L}  \\
%> & 1 - \left[(L+1)(|E| - \omega) + \omega L'\right] \cdot 2^{-L \epsilon_L}  \\
%\label{eqn:Thm_Proof_Bound_3}
%> & 1 - (L+1)|E|(1-\epsilon_L)2^{-L \epsilon_L}.
%\end{align}
\begin{align}
 \mathrm{Pr}(\mathrm{rank}(\mathbf{G}_s [\mathbf{F}_e]_{e \in \mathrm{In}(t)}\geq\omega L') \nonumber
> & 1 - (|E| - \omega) \cdot 2^{-L \epsilon_L+\log(L+1)} - \omega L' 2^{-\omega L \epsilon_L}  \\
> & 1 - \left[(L+1)(|E| - \omega) + \omega L'\right] \cdot 2^{-L \epsilon_L}  \\
\label{eqn:Thm_Proof_Bound_3}
> & 1 - (L+1)|E|(1-\epsilon_L)2^{-L \epsilon_L}.
\end{align}

By taking a union bound on (\ref{eqn:Thm_Proof_Bound_3}) for all receivers, the desired lower bound for the probability of the randomly constructed circular-shift linear code to be an $(L', L)$-fractional linear solution can be obtained.
\end{proof}
\end{theorem}

As a result, for an arbitrary multicast network, the probability for random circular-shift LNC to yield an asymptotic linear solution tends to $1$ with block length $L$ increasing to infinity. %
One may notice that in the work of \cite{Khojastepour10}, it was also proved that on a multicast network, the success probability of randomly generating an $(L-1, L)$ circular-shift linear solution (of degree 1) is lower bounded by $(1 - |T|/L)^{\sum_{v:\mathrm{node}} |\mathrm{In}(v)||\mathrm{Out}(v)|}$, %
the form of which is same as the classical lower bound $(1-|T|/2^L)^{|E|}$ obtained in \cite{Ho_Random_06} for the success probability of randomly generating a scalar linear solution over GF($2^L$). %
Compared with the one obtained in \cite{Khojastepour10}, when $L$ tends to infinity, the lower bound obtained in Theorem \ref{thm:asymptotic_solvable} converges to $1$ much faster for $L$ appears as an exponent parameter instead of as a denominator parameter. In addition, the rate $L'/L$ of the random code considered in Theorem \ref{thm:asymptotic_solvable} converges faster to $1$ compared with the rate $(L-1)/L$ of the random code considered in \cite{Khojastepour10}, too.

Moreover, circular-shift LNC of degree $1$ can be regarded as a special class of \emph{permutation-based} LNC schemes studied in \cite{Jaggi06}, in which the local encoding kernels are chosen from $L!$ permutation matrices of size $L$ as well as the $L\times L$ zero matrix $\mathbf{0}$. The bound in Theorem \ref{thm:asymptotic_solvable} is essentially the same as the lower bound obtained in \cite{Jaggi06} for the probability of a randomly constructed permutation-based linear code to be a linear solution. This connection is particular interesting because the coding operations provided by circular-shifts are much fewer than by permutations. Thus, the asymptotic linear solvability characterization in Theorem \ref{thm:asymptotic_solvable} is stronger than the results in \cite{Jaggi06}. %
We would remark here that to the best of our knowledge, the known analyses for random linear coding concentrate on special types of vector LNC, such as the scalar, the permutation-based, as well as the circular-shift LNC. There is not any more general lower bound on the success probability of randomly generating an $L$-dimensional vector linear solution with local encoding kernels selected from an arbitrary matrix of size $L$.

\subsection{Circular-shift LNC vs Permutation-based LNC}
In the previous subsection, we showed that the circular-shift LNC and permutation-based LNC essentially share the same lower bound obtained in Theorem \ref{thm:asymptotic_solvable} on the success probability of yielding an asymptotic linear solution. However, only when the block length $L$ is sufficiently long, the bound can start yielding a positive value. Therefore, it does not shed light on the asymptotic behavior for shorter block lengths. We next attempt to numerically analyze the success probability of randomly generating a circular-shift and a permutation-based linear solution of the same rate $r = L'/L = 15/16$ on the ($4,2$)-Combination Network, as shown in Table \ref{Table_probability_random_Coding}. It can be seen that even though the success probability for permutation-based LNC converges faster than the one for circular-shift LNC, for moderate block length $L = 128$, the success probabilities for both have no big difference and are very close to $1$.

\begin{table}[t]
\caption{success probability of randomly generating an $(L', L)$-fractional linear solution for the $(4, 2)$-Combination Network}
\label{Table_probability_random_Coding}
\begin{center}
\begin{tabular}{|c|c|c|}

  \hline
  ($L',L$) & Circular-shift & Permutation \\
  \hline
  ($15,16$) & 0.1055  & 0.0168                                            \\
  \hline
  ($30,32$) & 0.5894  & 0.3358                                            \\
  \hline
  ($60,64$) & 0.7031  & 0.9349                                            \\
  \hline
  ($120,128$) & 0.9996  & 0.9998                                            \\
  \hline
\end{tabular}
\end{center}
\end{table}

Though permutation-based LNC can be regarded as a generalization of circular-shift LNC (of degree 1), the above numerical result indicates that the much more local encoding kernel candidates it brings in ($L!$ vs $L+1$) do not obviously help increase the success probability of randomly constructing a solution. In addition, as to be shown in the next proposition, for both the $(n,2)$-Combination Network and the Swirl Network, which do not have an $(L, L)$ circular-shift linear solution for any $L$ as proved in Proposition \ref{prop:Insufficiency_Circular_Shift}, permutation-based LNC is insufficient to achieve their respective exact multicast capacity either.

\begin{proposition}
\label{prop:Insufficiency_Permutation}
For $n\geq 4$, neither the $(n, 2)$-Combination Network depicted in Fig. \ref{Fig:Combination_Network} nor the Swirl Network depicted in Fig. \ref{Fig:Swirl_Network} with parameter $\omega = n$ has an $L$-dimensional vector linear solution over GF($2$) with local encoding kernels chosen from the $L!$ possible permutation matrices of size $L$ and the $L\times L$ zero matrix $\mathbf{0}$, for any block length $L$.
\begin{proof}
See Appendix-\ref{Appendix:Prop_Proof_Permutation_Insuf}.
\end{proof}
\end{proposition}

It turns out that for multicast LNC, compared with permutation operations, circular-shifts do not lose much in terms of linear solvability, while they have much less implementation complexity.

\subsection{Overhead Analysis}
In the practical implementation of random LNC, every packet transmitted along the network usually consists of a batch of data units (See, e.g., \cite{Chou}). All data units belong to the same alphabet and all data units in the same packet correspond to the same global encoding kernel. %
When random LNC is applied to multicast networks, since the network topology is fixed, an initialization process can be conducted before the packet transmission so that every receiver can obtain the necessary information of global encoding kernels for decoding. However, in some other application scenarios of random LNC, such as the Peer-to-Peer networks (See, e.g., the review article \cite{Random_p2p}) and the Mobile Ad hoc Networks (MANETs) (See, e.g., \cite{P_Coding}), the network topology is always dynamic. %
It turns out that the global encoding kernel for a packet will be dynamically updated to indicate how the packet is linearly formed from the source packets, so its information must be stored as part of the packet header.

For a scalar linear code over GF($2^L$), as the global encoding kernels are $\omega$-dimensional vectors over GF($2^L$), the overhead to store the information of a global encoding kernel is theoretically $\omega L$ bits. On the other hand, for random vector LNC, under the same block length $L$, the global encoding kernel becomes an $\omega L\times L$ matrix over GF($2$) and thus the overhead to store the corresponding information theoretically extends to $\omega L^2$ bits. %
The next proposition considers the cases for random circular-shift LNC (of degree 1) and random permutation-based LNC, where the local encoding kernels are respectively randomly chosen from $\mathcal{C}_1 = \{\mathbf{0}, \mathbf{I}_L, \mathbf{C}_L, \ldots, \mathbf{C}^{L-1}_L\}$ and $L\times L$ permutation matrices.

\begin{table}[t]
\caption{Overheads of random LNC schemes under alphabet size $2^L$}
\label{Table_Comparison}
\begin{center}
\begin{tabular}{|c|c|}

  \hline
  Schemes & Overheads \\
  \hline
  Scalar LNC  &  $\omega L$ bits                                            \\
  \hline
  Circular-Shift LNC & $\omega L$ bits                                      \\
  \hline
  Permutation-based LNC & $\Omega(\omega L\log_{2}L)$                           \\                              \hline
  Vector LNC & $\omega L^2$ bits                                      \\
  \hline
\end{tabular}
\end{center}
\end{table}

\begin{proposition}
Under the same block length $L$, for a random circular-shift linear code and a random permutation-based linear code, the overheads to store the global encoding kernel information are $\omega L$ and $\Omega(\omega L \log_2 L)$ bits, respectively.

\begin{proof}
Recall that $[\mathbf{F}_e]_{e\in \mathrm{out}(s)} = \mathbf{I}_{\omega L}$, and for an outgoing edge $e$ from a non-source node $v$, the global encoding kernel $\mathbf{F}_e$ can be expressed as $\mathbf{F}_e = \sum_{d\in \mathrm{In}(v)} \mathbf{F}_d\mathbf{K}_{d,e}$. Then, when $\mathbf{F}_e$ is regarded as an $\omega$-dimensional vector with each component being an $L \times L$ matrix, each of these $\omega$ matrices can be recursively written as a function of local encoding kernels, which are randomly chosen from $\mathcal{C} = \{\mathbf{0}, \mathbf{I}, \mathbf{C}_L, \ldots, \mathbf{C}^{L-1}_L\}$. As $\mathcal{C}$ is closed under multiplication by elements in $\mathcal{C}$, each of the $\omega$ components in $\mathbf{F}_e$ is a summation of some matrices in $\mathcal{C}$. Thus, the number of possible $L \times L$ matrices to appear in each component of $\mathbf{F}_e$ is£º
\begin{equation}
\setlength{\arraycolsep}{2.0pt}
\renewcommand{\arraystretch}{0.6}
\left(\begin{matrix} L \\ 0 \end{matrix}\right) + \left(\begin{matrix} L \\ 1 \end{matrix} \right) + \ldots + \left(\begin{matrix} L \\ L \end{matrix} \right) = 2^L,
\end{equation}
which can be represented by $L$ bits. In all, the total number of bits required to store the information of $\mathbf{F}_e$ is $\omega L$.

For an $L$-dimensional permutation-based linear code, the number of local encoding kernel candidates is $L! = \Omega(\omega L\log_{2}L)$. As the number of possibilities for every block entry in a global encoding kernel $\mathbf{F}_e$ is at least the number of local encoding kernels, the overhead to store the information of $\mathbf{F}_e$ is $\Omega(\omega L\log_{2}L)$ bits.
\end{proof}
\end{proposition}

Table \ref{Table_Comparison} summarizes the required overheads for global encoding kernels among the aforementioned four types linear network coding schemes. The table shows that under the same alphabet size, the overhead required by random circular-shift LNC is as small as that required by conventional scalar LNC, and is much smaller than that of permutation-based LNC and vector LNC. The results established in this section show that circular-shift LNC also has advantages of shorter overheads for random coding and suggest a new direction of practical implementation of LNC using efficient, randomized circular-shift operations.

\section{Concluding Remarks}
In this work, after formulating circular-shift linear network coding (LNC) as a special type of vector LNC, we established an intrinsic connection between circular-shift and scalar LNC, for a \emph{general} network, so that the construction of a circular-shift linear solution with 1 bit redundancy is reduced to the construction of a scalar linear solution. The results subsequently obtained for multicast networks theoretically suggested the potential of circular-shift LNC to be deployed with lower implementation complexities in both deterministic and randomized manners, compared with the conventional scalar LNC and permutation-based LNC. In addition, they provided a method to efficiently construct a BASIC functional regenerating code for a distributed storage system proposed in \cite{HouShum16}.  %

With the aim to investigate LNC schemes with lower encoding and decoding complexities, the present paper focuses on the study of circular-shift LNC over GF($2$). An extension of the present work to GF($p$) with an odd prime $p$ is left as future work. %
In addition, whether every multicast network is asymptotically circular-shift linearly solvable remains open and it deserves further investigation. %
From a practical point of view, another important future work is to make a hardware-implemented experimental comparison of the encoding and decoding complexities between scalar and circular-shift LNC. %

\appendix
\subsection{Proof of Lemma \ref{lemma:cyclic_matrix_decomposition}}
\label{Appendix:Lemma_cyclic_matrix_decomposition_proof}
First note that the $i^{th}$ row in $\mathbf{V}_{L}$ times the $j^{th}$ column in $\mathbf{V}_{L}^{-1}$ ($0 \leq i, j \leq L-1$) equals to $\sum_{i'=0}^{L-1} \alpha^{i'(i-j)}$. Since $\alpha$ is a primitive $L^{th}$ root of unity, $\alpha^{i'}$ is a root of $x^L - 1$ and not equal to 1 for all $1 \leq i' \leq L-1$. In addition, since $x^L - 1 = (x-1)(x^{L-1} + \ldots + 1)$, $\sum_{i'=0}^{L-1}\alpha^{i'(i-j)} = 0$ when $i \neq j$. Furthermore, when $i = j$, $\sum_{i'=0}^{L-1} \alpha^{i'(i-j)} = 1$ for summation of $1$ by (odd) $L$ times is still equal to 1 over GF(2). In sum, $\mathbf{V}_L\mathbf{V}_L^{-1} = \mathbf{I}_L$.

Next, note that
\begin{equation}
\setlength{\arraycolsep}{3.0pt}
\renewcommand{\arraystretch}{0.7}
\mathbf{V}_{L}\cdot\mathbf{\Lambda}_\alpha =
\left[
\begin{matrix}
1 & \alpha & \ldots & \alpha^{L-1} \\
\vdots & \vdots & \ldots & \vdots \\
1 & \alpha^{L-1} & \ldots & \alpha^{(L-1)(L-1)}\\
1 & 1 & \ldots & 1
\end{matrix}
\right]
= \mathbf{C}_L \cdot \mathbf{V}_{L}.
\end{equation}
As a result,
\begin{equation} \mathbf{V}_{L}\cdot\mathbf{\Lambda}_\alpha\cdot\mathbf{V}_L^{-1}
= \mathbf{C}_L \cdot \mathbf{V}_{L}\cdot\mathbf{V}_L^{-1} = \mathbf{C}_L,
\end{equation}
and thus (\ref{eqn:cyclic_matrix_decomposition}) holds.

\subsection{Proof of Lemma \ref{lemma:extension_field}}
\label{Appendix:Lemma_Extension_Field_Proof}

\begin{enumerate}[a)]
\item As $0 = \alpha^L + 1 = (\alpha+1)(\alpha^{L-1} + \ldots + \alpha + 1)$ and $\alpha \neq 0$, $f(\alpha) = 0$. Consequently, $f(\alpha^{2^j}) = f(\alpha)^{2^j} = 0$ for all $j \geq 0$. As the multiplicative order of $2$ modulo $L$ is $L-1$, $\alpha, \alpha^2, \ldots, \alpha^{2^{L-2}}$ are distinct elements, and thus constitute the $L - 1$ roots of $f(x)$. This implies that $f(x)$ is irreducible over GF($2$), so $\alpha \in \mathrm{GF}(2^{L-1})$.

\item Because $f(x)$ is irreducible over GF($2$) and $f(\alpha) = 0$, $\{1, \alpha, \ldots, \alpha^{L-2}\}$ is a basis of GF($2^{L-1}$) over GF($2$). Thus, every element $k \in \mathrm{GF}(2^{L-1})$ can be uniquely written as $a_0 + a_1\alpha + \ldots + a_{L-2}\alpha^{L-2}$ with the binary coefficients $a_j$, $0 \leq j \leq L-2$. Additionally set $a_{L-1}$ to be $0$. If the number of nonzero coefficients $a_j$ is no larger than $\frac{L-1}{2}$, then $g(x) = a_{L-1}x^{L-1} + \ldots + a_1x + a_0$ is a polynomial in the form of (\ref{eqn:field-element-unique-expression-polynomial}) with $g(\alpha) = k$. Otherwise, set $a_j' = 1\oplus a_j$ for all $0 \leq j \leq L-1$. In this way, $g(x) = a_{L-1}'x^{L-1} + \ldots + a_1'x + a_0'$ is a polynomial in the form of (\ref{eqn:field-element-unique-expression-polynomial}) with at most $\frac{L-1}{2}$ nonzero terms and $g(\alpha) = k$. As there are in total $2^{L-1}$ polynomials over GF($2$) in the form of (\ref{eqn:field-element-unique-expression-polynomial}) with at most $\frac{L-1}{2}$ nonzero terms, each of the $2^{L-1}$ polynomials has been associated with a distinct element in GF($2^{L-1}$).

\item As the multiplicative order of $2$ modulo $L$ is $L-1$, for each $1 \leq j \leq L-1$, there exists $i \geq 1$ such that $\alpha^j = \alpha^{2^i}$. Thus, when $g_1(\alpha^{k_1}) = g_2(\alpha^{k_2})$,
    \begin{equation}
    g_1(\alpha^{jk_1}) = g_1(\alpha^{2^ik_1}) = g_1(\alpha^{k_1})^{2^i} = g_2(\alpha^{k_2})^{2^i} = g_2(\alpha^{2^ik_2}) = g_2(\alpha^{jk_2}).
    \end{equation}
\end{enumerate}

\subsection{Proof of Theorem \ref{thm:deterministic_construction}}
\label{Appendix:Main-Thm-Proof}
For every edge $e \in E$, denote by $\mathbf{F}_e$ and $\mathbf{f}_e$ the global encoding kernels of the considered $(L-1, L)$-fractional linear code $(\mathbf{K}_{d,e})$ over GF($2$) and scalar linear solution $(g_{d,e}(\alpha))$ over $\mathrm{GF}(2^{L-1})$, respectively. For brevity, write $E_S = \mathrm{Out}(S)$ and $E_{S_t} = \mathrm{Out}(S_t)$.

Consider an arbitrary receiver $t$. Denote by $\mathbf{B}_t(x)$ the $(|E|-\omega)\times |\mathrm{In}(t)|$ index matrix of which the unique nonzero entry $x$ in every column corresponds to an edge in $\mathrm{In}(t)$. Thus, $[\mathbf{f}_e]_{e\in \mathrm{In}(t)} = [\mathbf{f}_e]_{e \notin E_S}\mathbf{B}_t(1)$ and $[\mathbf{F}_e]_{e\in \mathrm{In}(t)} = [\mathbf{F}_e]_{e\notin E_S}\mathbf{B}_t(\mathbf{I}_L)$.
Following the classic algebraic framework of scalar LNC for acyclic multicast networks in \cite{KoetterMedard03}, %
the global encoding kernels of the scalar linear code $(g_{d,e}(\alpha))$ for edges into $t$ can be expressed as
\begin{equation}
\label{Eqn:fe_In_t_expression}
[\mathbf{f}_e]_{e \in \mathrm{In}(t)}
= [g_{d,e}(\alpha)]_{d\in E_S, e\notin E_S}\cdot(\mathbf{I}_{|E|-\omega} - [g_{d,e}(\alpha)]_{d,e\notin E_S})^{-1}\cdot\mathbf{B}_t(1).
\end{equation}%
Note that (\ref{Eqn:fe_In_t_expression}) is essentially the same as the formula in Theorem 3 of \cite{KoetterMedard03}.
Write the matrix $[\mathbf{f}_e]_{e\in \mathrm{In}(t)}$ over GF($2^{L-1}$) as $\mathbf{M}(\alpha)$, where $\mathbf{M}(x)$ is the matrix over GF($2$)$[x]$ with every entry to be a polynomial of at most $\frac{L-1}{2}$ nonzero terms. Thus,
\begin{equation}
\label{eqn:Scalar_Solution_Matrix_Product}
\mathbf{M}(\alpha)\mathbf{D}_t(\alpha) = [\mathbf{U}_e^1]_{e\in E_{S_t}}.
\end{equation}

Now consider the $(L-1, L)$-fractional code with the local encoding kernels $\mathbf{K}_{d, e} = g_{d,e}(\mathbf{C}_L)$. According to the framework of vector LNC \cite{Ebrahimi},
\begin{align}
[\mathbf{F}_e]_{e \in \mathrm{In}(t)} &= [\mathbf{K}_{d, e}]_{d\in E_s, e\notin E_s}\cdot \left(\mathbf{I}_{(|E|-\omega)L} + [\mathbf{K}_{d, e}]_{d,e\notin E_s} + \ldots + [\mathbf{K}_{d, e}]_{d,e\notin E_s}^{|E|}\right)\cdot\mathbf{B}_t(\mathbf{I}_L) \\
& = [\mathbf{K}_{d, e}]_{d\in E_s, e\notin E_s}\cdot\left(\mathbf{I}_{(|E|-\omega)L} - [\mathbf{K}_{d, e}]_{d,e\notin E_s}\right)^{-1}\cdot\mathbf{B}_t(\mathbf{I}_L)
\end{align}
By Lemma \ref{lemma:cyclic_matrix_decomposition},
$\mathbf{K}_{d, e} = g_{d,e}(\mathbf{C}_L) = \mathbf{V}_L \cdot g_{d,e}(\mathbf{\Lambda}_\alpha) \cdot \mathbf{V}_L^{-1}.
$ %
Thus,
\begin{equation}
[\mathbf{K}_{d,e}]_{d\in E_s, e\notin E_s}
=(\mathbf{I}_\omega\otimes\mathbf{V}_L)\cdot[g_{d,e}(\mathbf{\Lambda}_\alpha)]_{d\in E_s, e\notin E_s}\cdot (\mathbf{I}_{|E|-\omega}\otimes\mathbf{V}_L^{-1})
\end{equation}
\begin{equation}
[\mathbf{K}_{d,e}]_{d,e\notin E_s}^j
=(\mathbf{I}_{|E|-\omega}\otimes\mathbf{V}_L)\cdot[g_{d,e}(\mathbf{\Lambda}_\alpha)]_{d,e\notin E_s}^j \cdot
(\mathbf{I}_{|E|-\omega}\otimes\mathbf{V}_L^{-1})~~~~~\forall 1 \leq j \leq |E|
\end{equation}
In addition, note that
\begin{equation}
\mathbf{B}_t(\mathbf{I}_L) = (\mathbf{I}_{|E|-\omega}\otimes\mathbf{V}_L) \cdot \mathbf{B}_t(\mathbf{I}_L)\cdot (\mathbf{I}_{|\mathrm{In}(t)|}\otimes\mathbf{V}_L^{-1}).
\end{equation}
Consequently, $[\mathbf{F}_e]_{e \in \mathrm{In}(t)}
= (\mathbf{I}_\omega\otimes\mathbf{V}_L) \cdot \tilde{\mathbf{M}}\cdot (\mathbf{I}_{|\mathrm{In}(t)|}\otimes\mathbf{V}_L^{-1})$,
where $\tilde{\mathbf{M}}$ represents the $\omega L \times |\mathrm{In}(t)|L$ matrix
\begin{align}
%&\tilde{\mathbf{M}} \\
&[g_{d,e}(\mathbf{\Lambda}_\alpha)]_{d\in E_s, e\notin E_s}\cdot
\left(\mathbf{I}_{(|E|-\omega)L} + [g_{d,e}(\mathbf{\Lambda}_\alpha)]_{d,e\notin E_s}+ \ldots + [g_{d,e}(\mathbf{\Lambda}_\alpha)]_{d,e\notin E_s}^{|E|}\right)\cdot \mathbf{B}_t(\mathbf{I}_L) \\
= &[g_{d,e}(\mathbf{\Lambda}_\alpha)]_{d\in E_s, e\notin E_s}\cdot
\left(\mathbf{I}_{(|E|-\omega)L} - [g_{d,e}(\mathbf{\Lambda}_\alpha)]_{d,e\notin E_s}\right)^{-1}\cdot \mathbf{B}_t(\mathbf{I}_L)
\end{align}
In the decoding matrix $\mathbf{D}_t(\mathbf{C}_L)\cdot(\mathbf{I}_{\omega_t}\otimes\tilde{\mathbf{I}}_L)$, note that
\begin{equation}
\mathbf{D}_t(\mathbf{C}_L) = (\mathbf{I}_{|\mathrm{In}(t)|}\otimes\mathbf{V}_L)\cdot \mathbf{D}_t(\mathbf{\Lambda}_\alpha)\cdot (\mathbf{I}_{\omega_t}\otimes\mathbf{V}_L^{-1}).
\end{equation} %
Thus,
\begin{equation}
\label{eqn:Thm_Proof_1}
[\mathbf{F}_e]_{e\in \mathrm{In}(t)}\cdot\mathbf{D}_t(\mathbf{C}_L) = (\mathbf{I}_\omega\otimes\mathbf{V}_L) \cdot \tilde{\mathbf{M}}\cdot \mathbf{D}_t(\mathbf{\Lambda}_\alpha) \cdot (\mathbf{I}_{\omega_t}\otimes\mathbf{V}_L^{-1}).
\end{equation}
Observe that both $\tilde{\mathbf{M}}$ and $\mathbf{D}_t(\mathbf{\Lambda}_\alpha)$ can be respectively regarded as an $\omega \times |\mathrm{In}(t)|$ and an $|\mathrm{In}(t)| \times \omega_t$ block matrix, and every block entry is an $L \times L$ diagonal matrix. %
Hence, $\tilde{\mathbf{M}}\cdot\mathbf{D}_t(\mathbf{\Lambda}_\alpha)$ is an $\omega \times \omega_t$ block matrix with every block entry being an $L\times L$ diagonal matrix.
Define an $\omega L \times \omega L$ permutation matrix $\mathbf{P}_j$ (over GF($2$)) as follows. It is an $L \times \omega$ block matrix
$\left[\begin{smallmatrix} \mathbf{J}_{1,1} & \ldots & \mathbf{J}_{1, \omega} \\ \vdots & \ddots \vdots \\ \mathbf{J}_{L,1} & \ldots & \mathbf{J}_{L, \omega} \end{smallmatrix}\right]$
in which the only nonzero entry in the $\omega \times L$ matrix $\mathbf{J}_{i,j}$ is in row $j$ and column $i$.
Rearrange the rows and columns in $\tilde{\mathbf{M}}\cdot\mathbf{D}_t(\mathbf{\Lambda}_\alpha)$ by respectively left-multiplying $\mathbf{P}_\omega$ and right-multiplying $\mathbf{P}_{\omega_t}^T$ to it. %
In this way, $\mathbf{P}_\omega\left(\tilde{\mathbf{M}}\cdot\mathbf{D}_t(\mathbf{\Lambda}_\alpha)\right)\mathbf{P}_{\omega_t}^T$ becomes an $L \times L$ block diagonal entry. %
The $j^{th}$ diagonal block entry, $0 \leq j \leq L-1$, in it is an $\omega \times \omega_t$ matrix
\begin{equation}
\label{eqn_3_2}
[g_{d,e}(\alpha^j)]_{d\in E_s, e\notin E_s}\cdot \left(\mathbf{I}_{(|E|-\omega)L} - [g_{d,e}(\alpha^j)]_{d,e\notin E_s}\right)^{-1}\cdot \mathbf{B}_t(1) \cdot \mathbf{D}_t(\alpha^j) = \mathbf{M}(\alpha^j)\cdot \mathbf{D}_t(\alpha^j),
\end{equation}
where the equality holds because of the definition of $\mathbf{M}(\alpha)$ and Lemma \ref{lemma:extension_field}.c). In total,
\begin{equation}
\setlength{\arraycolsep}{2.0pt}
\renewcommand{\arraystretch}{0.6}
\mathbf{P}_\omega \left(\tilde{\mathbf{M}}\cdot\mathbf{D}_t(\mathbf{\Lambda}_\alpha)\right)\mathbf{P}_{\omega_t}^T = \left[\begin{matrix}
\mathbf{M}(1)\mathbf{D}_t(1) & \mathbf{0} & \ldots & \mathbf{0} \\
\mathbf{0} & \mathbf{M}(\alpha)\mathbf{D}_t(\alpha) & \ddots & \vdots \\
\vdots & \ddots & \ddots & \mathbf{0} \\
\mathbf{0} & \ldots & \mathbf{0} & \mathbf{M}(\alpha^{L-1})\mathbf{D}_t(\alpha^{L-1})
\end{matrix}\right].
\end{equation}
By (\ref{eqn:Scalar_Solution_Matrix_Product}), $\mathbf{M}(\alpha)\mathbf{D}_t(\alpha) = [\mathbf{U}_e^1]_{e\in E_{S_t}}$. As a consequence of Lemma \ref{lemma:extension_field}.c),
\begin{equation}
\mathbf{M}(\alpha^j)\mathbf{D}_t(\alpha^j) = [\mathbf{U}_e^1]_{e\in E_{S_t}}~~~\forall 1 \leq j \leq L-1.
\end{equation}
In addition, write $\mathbf{M}(1)\mathbf{D}_t(1) = \left[
\begin{smallmatrix} a_{11} & \ldots & a_{1\omega_t} \\ \vdots & \ddots & \vdots \\ a_{\omega1} & \ldots & a_{\omega\omega_t} \end{smallmatrix} \right]$. Note that the entries $a_{ij}$ belong to GF($2$). Then,
\begin{equation}
\label{eqn:Thm_Proof_2}
\tilde{\mathbf{M}}\cdot\mathbf{D}_t(\mathbf{\Lambda}_\alpha)
=
\setlength{\arraycolsep}{2.0pt}
\renewcommand{\arraystretch}{0.6}
\left[\begin{matrix}
\begin{smallmatrix}
a_{11} & \mathbf{0} \\ \mathbf{0} &  \mathbf{J}_{11}
\end{smallmatrix}
& \ldots &
\begin{smallmatrix}
a_{1\omega_t} & \mathbf{0} \\ \mathbf{0} & \mathbf{J}_{1\omega_t}
\end{smallmatrix} \\
\vdots & \ddots & \vdots \\
\begin{smallmatrix}
a_{\omega 1} & \mathbf{0} \\ \mathbf{0} & \mathbf{J}_{\omega 1}
\end{smallmatrix}
& \ldots &
\begin{smallmatrix}
a_{\omega\omega_t} & \mathbf{0} \\ \mathbf{0} & \mathbf{J}_{\omega\omega_t}
\end{smallmatrix}
\end{matrix}\right],
\end{equation}
where $\mathbf{J}_{i,j}$, $1\leq i \leq \omega$, $1\leq j\leq \omega_t$, is set to $\mathbf{I}_{L-1}$ if the $(i,j)^{th}$ entry in $[\mathbf{U}_e^1]_{e\in E_{S_t}}$ is equal to 1, and set to the $(L-1)\times (L-1)$ zero matrix otherwise.
Let $\hat{\mathbf{I}}_L$ denote the $L \times L$ matrix which is identical to $\mathbf{I}_L$ except for the $(1, 1)^{st}$ entry equal to $0$, and $\mathbf{1}_L$ denote the $L\times L$ matrix with all entries equal to $1$. It can be readily checked that
\begin{equation}
\label{eqn:Thm_Proof_3}
\mathbf{V}_L \cdot \hat{\mathbf{I}}_L \cdot (\mathbf{1}_L + \mathbf{V}_L^{-1})\cdot \tilde{\mathbf{I}}_L = \tilde{\mathbf{I}}_L.
\end{equation}
Based on (\ref{eqn:Thm_Proof_1}), (\ref{eqn:Thm_Proof_2}) and (\ref{eqn:Thm_Proof_3}), we have
\begin{align}
&[\mathbf{F}_e]_{e\in \mathrm{In}(t)}\cdot\mathbf{D}_t(\mathbf{C}_L)\cdot (\mathbf{I}_{\omega_t}\otimes\tilde{\mathbf{I}}_L) \\
= &(\mathbf{I}_\omega\otimes\mathbf{V}_L) \cdot \tilde{\mathbf{M}}\cdot \mathbf{D}_t(\mathbf{\Lambda}_\alpha) \cdot (\mathbf{I}_{\omega_t}\otimes\mathbf{V}_L^{-1})\cdot(\mathbf{I}_{\omega_t}\otimes\tilde{\mathbf{I}}_L) \\
= &(\mathbf{I}_\omega\otimes\mathbf{V}_L) \cdot \tilde{\mathbf{M}}\cdot \mathbf{D}_t(\mathbf{\Lambda}_\alpha) \cdot (\mathbf{I}_{\omega_t}\otimes(\hat{\mathbf{I}}_L \cdot (\mathbf{1}_L + \mathbf{V}_L^{-1})\cdot \tilde{\mathbf{I}}_L)) \\
= &(\mathbf{I}_\omega\otimes\mathbf{V}_L) \cdot
\setlength{\arraycolsep}{2.0pt}
\renewcommand{\arraystretch}{0.6}
\left[\begin{matrix}
\begin{smallmatrix}
a_{11} & \mathbf{0} \\ \mathbf{0} &  \mathbf{J}_{11}
\end{smallmatrix}
& \ldots &
\begin{smallmatrix}
a_{1\omega_t} & \mathbf{0} \\ \mathbf{0} & \mathbf{J}_{1\omega_t}
\end{smallmatrix} \\
\vdots & \ddots & \vdots \\
\begin{smallmatrix}
a_{\omega 1} & \mathbf{0} \\ \mathbf{0} & \mathbf{J}_{\omega 1}
\end{smallmatrix}
& \ldots &
\begin{smallmatrix}
a_{\omega\omega_t} & \mathbf{0} \\ \mathbf{0} & \mathbf{J}_{\omega\omega_t}
\end{smallmatrix}
\end{matrix}\right]
\cdot (\mathbf{I}_{\omega_t}\otimes(\hat{\mathbf{I}}_L \cdot (\mathbf{1}_L + \mathbf{V}_L^{-1})\cdot \tilde{\mathbf{I}}_L)) \\
= &(\mathbf{I}_\omega\otimes\mathbf{V}_L) \cdot ([\mathbf{U}_e^1]_{e\in E_{S_t}}\otimes\hat{\mathbf{I}}_L)\cdot (\mathbf{I}_{\omega_t}\otimes(\mathbf{1}_L + \mathbf{V}_L^{-1})\cdot \tilde{\mathbf{I}}_L) \\
= &[\mathbf{U}_e^1]_{e\in E_{S_t}} \otimes \tilde{\mathbf{I}}_L.
\end{align}
Finally, as for each $e \in \mathrm{Out}(S)$, the binary sequences transmitted on $e$ is $[0~\mathbf{m}_e']$, $G_S = \mathbf{I}_\omega \otimes [\mathbf{0}~~\mathbf{I}_{L-1}]$, \emph{i.e.},
\begin{equation}
[\mathbf{m}_e]_{e \in \mathrm{Out}(S)} = [\mathbf{m}_e']_{e \in \mathrm{Out}(S)}\cdot\left(\mathbf{I}_\omega \otimes [\mathbf{0}~~\mathbf{I}_{L-1}]\right).
\end{equation}
In summary,
\begin{align}
&\mathbf{G}_S \cdot[\mathbf{F}_e]_{e\in \mathrm{In}(t)}\cdot\mathbf{D}_t(\mathbf{C}_L)\cdot (\mathbf{I}_{\omega_t}\otimes\tilde{\mathbf{I}}_L) \\
= &(\mathbf{I}_\omega \otimes [\mathbf{0}~~\mathbf{I}_{L-1}]) \cdot ([\mathbf{U}_e^1]_{e\in E_{S_t}} \otimes \tilde{\mathbf{I}}_L) = \mathbf{U}_t \otimes \mathbf{I}_{L-1} = [\mathbf{U}_e^{L-1}]_{e\in E_{S_t}},
\end{align}
{\em i.e.}, receiver $t$ can recover $(L-1)$-dimensional source row vectors $\mathbf{m}_e'$, $e \in \mathrm{Out}(S_t)$ generated by sources in $S_t$ based on the decoding matrix $\mathbf{D}_t(\mathbf{C}_L)
\cdot(\mathbf{I}_{\omega_t}\otimes\tilde{\mathbf{I}}_L)$.

\subsection{Proof of Lemma \ref{lemma:L+K}}
\label{Appendix:Lemma_Proof}
For a fixed $L$-dimensional vector $\mathbf{v}$ over GF(2), the probability %of the random matrix $\mathbf{A}$ over $\{\mathbf{I}_L, \mathbf{C}, \ldots, \mathbf{C}^{L - 1}\}$
that $\mathbf{v}$ is in the null-space of $\mathbf{L}+\mathbf{K}$ is
%\begin{align}
%& Pr[(\mathbf{L}+\mathbf{K})\mathbf{v} = \mathbf{0}] \nonumber \\
%= & Pr[\mathbf{v}'=\mathbf{K}\mathbf{v}] \\
%= &
%\label{eqn:Prob_expression_Lv=Kv}
%\begin{cases}
%\cfrac{1}{\binom{L}{w_H(\mathbf{v})}},&
%w_H(\mathbf{v}')=w_H(\mathbf{v})     \\
%0,&\mathrm{otherwise}                           \\
%\end{cases}
%\end{align}
\begin{equation}
\label{eqn:Prob_expression_Lv=Kv}
\mathrm{Pr}((\mathbf{L}+\mathbf{K})\mathbf{v} = \mathbf{0})  \\
= \mathrm{Pr}(\mathbf{v}'=\mathbf{K}\mathbf{v})
=
\begin{cases}
\cfrac{1}{\binom{L}{w_H(\mathbf{v})}},&
w_H(\mathbf{v}')=w_H(\mathbf{v})     \\
0,&\mathrm{otherwise}                           \\
\end{cases}
\end{equation}
where $\mathbf{v}'=\mathbf{L}\mathbf{v}$, and $w_H(\cdot)$ stands for the Hamming weight of a vector. The reason for (\ref{eqn:Prob_expression_Lv=Kv}) to hold is as follows. First, note that since $\mathbf{K}$ acts as a random circular-shift operation on  $\mathbf{v}$, $\mathbf{v}'=\mathbf{K}\mathbf{v}$ only if $w_H(\mathbf{v}') = w_H(\mathbf{v})$.
%, and at the same time $\mathbf{v}$ and $\mathbf{L}\mathbf{v}$ have the same order of binary bits.
Next, when $\mathbf{K}$ is chosen from $\{\mathbf{0},\mathbf{I}_L, \mathbf{C}_L, \ldots, \mathbf{C}^{L - 1}_L\}$, there are $l\leq L$ vectors $\mathbf{v}'$ subject to $\mathbf{v}' = \mathbf{K}\mathbf{v}$. %
As it is possible that $\mathbf{C}^i \mathbf{v} = \mathbf{C}^j \mathbf{v}$ for some $0 \leq i < j \leq L -1$, $l$ can be strictly smaller than $L$.
For the $i^{th}$ possible vector $\mathbf{v}'$ subject to $\mathbf{v}' = \mathbf{K}\mathbf{v}$, let $t_i$ be the number of matrices $\mathbf{C}^i_L$, $0 \leq i \leq L - 1$ subject to $\mathbf{v}' = \mathbf{C}^i_L\mathbf{v}$. Apparently, $\sum_{i = 1}^l t_i = L$. Then,
\begin{align}
\mathrm{Pr}(\mathbf{L}\mathbf{v}=\mathbf{K}\mathbf{v}) = \sum_{i=1}^{l}\frac{1}{{\binom{L}{w_H(\mathbf{v})}}}\times \frac{t_i}{L}
=\frac{1}{L{\binom{L}{w_H(\mathbf{v})}}}\sum_{i=1}^{l}t_i
= \frac{L}{L{\binom{L}{w_H(\mathbf{v})}}}
= \frac{1}{\binom{L}{w_H(\mathbf{v})}}
\end{align}

Now let $\mathbf{v}$ be chosen uniformly and randomly from $L$-dimensional binary vectors. Then the probability that $\mathbf{v}$ is in the null-space of $\mathbf{L}+\mathbf{K}$ is
%\begin{equation}
\begin{align}
\label{eqn_Appendix_1}
\mathrm{Pr}((\mathbf{L+K})\mathbf{v}=\mathbf{0}) \leq \frac{1}{2^L}\sum_{\mathbf{v}}\frac{1}{\binom{L}{w_H(\mathbf{v})}}
=\frac{1}{2^L}\sum_{i=0}^{L}\binom{L}{w_H(\mathbf{v})}\frac{1}{\binom{L}{w_H(\mathbf{v})}}
=\frac{L+1}{2^L},
\end{align}
%\end{equation}
where the inequality in (\ref{eqn_Appendix_1}) holds due to the partitioning of the set of all $L$-dimensional binary vectors $\mathbf{v}$ into $L+1$ classes of different Hamming weights. Since there are $L+1$ random choices for $\mathbf{K}$ and $2^L$ random choices for $\mathbf{v}$, the number of ($\mathbf{v},\mathbf{K}$) pairs satisfying $\mathbf{L}\mathbf{v}=\mathbf{K}\mathbf{v}$ is bounded by
\begin{equation}
\label{eqn:Lemma_bound}
(L+1)\times 2^L\mathrm{Pr}((\mathbf{L+K})\mathbf{v}=\mathbf{0})\leq (L+1)2^L\times\frac{L+1}{2^L} = (L+1)^2.
\end{equation}
Let $k$ denote the number of choices for $\mathbf{K}$ such that
\begin{equation}
\label{eqn:Lemma_rank}
\mathrm{rank}(\mathbf{L}+\mathbf{K}) < L(1-\epsilon).
\end{equation}
For each $\mathbf{K}$ subject to (\ref{eqn:Lemma_rank}), the number of vectors $\mathbf{v}$ in the null space of $\mathbf{L} + \mathbf{K}$ is at least $2^{L(1-\epsilon)}$, i.e., the number of ($\mathbf{v},\mathbf{K}$) pairs satisfying $\mathbf{L}\mathbf{v}=\mathbf{K}\mathbf{v}$ is at least $2^{L(1-\epsilon)}$. Thus, as a consequence of (\ref{eqn:Lemma_bound}),
\begin{equation}
k < \frac{(L+1)^2}{2^{L(1-\epsilon)}}.
\end{equation}
Since there are $L$ possible choices for $\mathbf{K}$ in total, the desired probability is upper bounded by $[(L+1)^2/2^{L\epsilon}]/L+1=(L+1)/2^{L\epsilon}=2^{-L\epsilon+\log(L+1)}$.

\subsection{Justification of Bounds (\ref{eqn:Thm_Proof_Bound_0}) and (\ref{eqn:Thm_Proof_Bound_2})}
\label{Appendix:Bound_Justification}
In this appendix, we provide a detailed proof on obtaining the bounds (\ref{eqn:Thm_Proof_Bound_0}) and (\ref{eqn:Thm_Proof_Bound_2}). Adopt the same notations as in the proof sketch following Theorem \ref{thm:asymptotic_solvable}.

First we shall prove inequality (\ref{eqn:Thm_Proof_Bound_0}). Recall that in the $i^{th}$ round of the iterative process, $E_\omega'$ is formed from $E_\omega$ via substituting $e_{j}$ by $e_{i+\omega}$, where $(e_j, e_{i+\omega})$ forms an adjacent pair of edges. Let $\hat{\mathbf{F}}$ be any $\omega L \times K$ submatrix of $[\mathbf{F}_e]_{e\in E_\omega}$ with $\mathrm{rank}([\mathbf{F}_e]_{e\in E_\omega})=\mathrm{rank}(\hat{\mathbf{F}})=K$. Write $\hat{\mathbf{F}} = [\hat{\mathbf{F}}_1~\hat{\mathbf{F}}_0]$, where $\hat{\mathbf{F}}_0$, $\hat{\mathbf{F}}_1$ respectively consist of columns in $\mathbf{F}_{e_j}$ and $[\mathbf{F}_e]_{e\in E_\omega\backslash\{e_j\}}$ that are contained in $\hat{\mathbf{F}}$. %
Because $\mathrm{rank}([\mathbf{F}_e]_{e\in E_\omega\backslash\{e_j\}}) \geq \mathrm{rank}(\hat{\mathbf{F}}_1)$ and $[\mathbf{F}_e]_{e\in E_\omega'} = [[\mathbf{F}_e]_{e\in E_\omega\backslash\{e_j\}}~~\mathbf{F}_{e_{i + \omega}}]$,
\begin{align}
\mathrm{rank}([\mathbf{F}_e]_{e\in E_\omega}) - \mathrm{rank}([\mathbf{F}_e]_{e\in E_\omega'})
\leq \mathrm{rank}([\hat{\mathbf{F}}_1~\hat{\mathbf{F}}_0]) - \mathrm{rank}([\hat{\mathbf{F}}_1~\mathbf{F}_{e_{i+\omega}}]).
\end{align}
In order to prove the bound (\ref{eqn:Thm_Proof_Bound_0}) for general cases, it suffices to prove (\ref{eqn:Thm_Proof_Bound_0}) under the assumption that the columns in $\mathbf{F}_{e_{i+\omega}}$ are only linearly dependent on column vectors in $\hat{\mathbf{F}}$. %
Then, there must exist $L \times L$ matrices $\hat{\mathbf{L}}_1$, $\hat{\mathbf{L}}_2$, and a randomly generated cyclic permutation matrix $\mathbf{K}_{e_j,e_{i+\omega}}$ (the local encoding kernel for adjacent pair $(e_j, e_{i+\omega})$) such that
\begin{equation}
[\hat{\mathbf{F}}_1~\mathbf{F}_{e_{i+\omega}}] = [\hat{\mathbf{F}}_1~\hat{\mathbf{F}}_0]
\begin{bmatrix}  \mathbf{I}_{K-\hat{L}} & \hat{\mathbf{L}}_1 \\
\mathbf{0} & \hat{\mathbf{L}}_2+\mathbf{K}_{e_{j},e_{i+\omega}}
\end{bmatrix},
\end{equation}
where $\hat{L}$ refers to the number of columns in $\hat{\mathbf{F}}_0$.
Subsequently,
\begin{align}
&\mathrm{Pr}(\mathrm{rank}([\mathbf{F}_e]_{e\in E_\omega}) - \mathrm{rank}([\mathbf{F}_e]_{e\in E_\omega'}) > L\epsilon_L) \nonumber \\
\leq &\mathrm{Pr}(\mathrm{rank}([\hat{\mathbf{F}}_1~\hat{\mathbf{F}}_0]) - \mathrm{rank}([\hat{\mathbf{F}}_1~\mathbf{F}_{e_{i+\omega}}]) > L\epsilon_L) \\
= &\mathrm{Pr}(\mathrm{rank}(\hat{\mathbf{L}}_2+\mathbf{K}_{e_{j},e_{i+\omega}}) < L - L\epsilon_L) \leq 2^{-L \epsilon_L+\log(L+1)}
\end{align}
where the last inequality is a direct consequence of Lemma \ref{lemma:L+K}. The bound (\ref{eqn:Thm_Proof_Bound_0}) is thus established.

Next, we shall prove inequality (\ref{eqn:Thm_Proof_Bound_2}). Assume that $r = \omega L - L\epsilon_L (|E| - \omega)$. Under this condition, the number of choices for the $\omega L' \times \omega L$ binary  matrix $\mathbf{G}_s$ satisfying $\mathrm{rank}(\mathbf{G}_s [\mathbf{F_e}]_{e \in \mathrm{In}(t)}) \geq \omega L'$ is equal to
\begin{equation}
(2^{\omega L}-2^{\omega L - r})(2^{\omega L}-2^{\omega L - r+1})\ldots (2^{\omega L}-2^{\omega L - r +\omega L'-1}).
\end{equation}
As $\mathbf{G}_s$ is uniformly and randomly chosen from all $2^{(\omega L')(\omega L)}$ possible $\omega L' \times \omega L$ binary matrices,
\begin{align}
\mathrm{Pr}(\mathrm{rank}&(\mathbf{G}_s [\mathbf{F}_e]_{e \in \mathrm{In}(t)}) \geq \omega L'~|~ \mathrm{rank}([\mathbf{F}_e]_{e \in \mathrm{In}(t)}) \geq r) \nonumber \\
&=\frac{(2^{\omega L}-2^{\omega L - r})\ldots (2^{\omega L}-2^{\omega L - r +\omega L'-1})}{2^{(\omega L)(\omega L')}} \\
&=(1 - 2^{-r})(1 - 2^{-r+1})\ldots(1-2^{-r+\omega L' - 1}) \\
&> (1-2^{-r+\omega L' - 1})^{\omega L'}  \\
&= (1-2^{-\omega L\epsilon_L - 1})^{\omega L'} \\
&> 1-\omega L'2^{-\omega L\epsilon_L}.
\end{align}
%Herein the second last inequality  holds due to $r \leq |E|L\epsilon_L < (|E|+1)L\epsilon_L \leq \omega L - \omega L'$, and the last inequality holds for sufficiently large $L$.
Inequality (\ref{eqn:Thm_Proof_Bound_2}) has thus been established.

\subsection{Proof of Proposition \ref{prop:Insufficiency_Permutation}}
\label{Appendix:Prop_Proof_Permutation_Insuf}
Same as in the proof of Proposition \ref{prop:Insufficiency_Circular_Shift}, we start the proof from the following necessary condition for both the $(n, 2)$-Combination Network and the Swirl Network with $|\mathrm{Out}(s)| = n$ to be $L$-dimensional vector linearly solvable over GF($2$): there are two $L \times L$ invertible matrices $\mathbf{A}_i, \mathbf{A}_j$ over GF($2$) such that
%It has been proved in \cite{Sun_TCom16} that the $(n, 2)$-Combination network is {\color{red} $L$-dimensional }vector linearly solvable over GF($2$) if and only if there are $L \times L$ invertible matrices $\mathbf{A}_1, \cdots, \mathbf{A}_{n-2}$ over GF($2$) such that
\begin{equation}
\label{eqn:combination_network_vector_LNC_1}
\mathrm{rank}(\mathbf{A}_i -\mathbf{A}_j) = L
\end{equation}
%Similarly, we can show that the $(n, 2)$-Combination Network has an {\color{red} $L$-dimensional } vector linear solution over GF($2$) with local encoding kernels chosen from permutation matrices of size $L$ and $\mathbf{0}$ if and only if there are $L \times L$ matrices $\mathbf{A}_1, \cdots, \mathbf{A}_{n-2}$ each of which is chosen from permutation matrices of size $L$ such that
%\begin{equation}
%\label{eqn:combination_network_vector_LNC_2}
%rank(\mathbf{A}_i -\mathbf{A}_j) = L,~~\forall 1 \leq i < j \leq n-2.
%\end{equation}

It suffices to show that $\mathrm{rank}(\mathbf{A}_{i}-\mathbf{A}_{j}) < L$ for two arbitrary permutation matrices of size $L$. First note that each of $\mathbf{A}_i$ and $\mathbf{A}_j$ has exactly one non-zero entry in every row and every column. In the case that $\mathbf{A}_{i}$ and $\mathbf{A}_{j}$ have a non-zero entry at a same position, $\mathbf{A}_{i}-\mathbf{A}_{j}$ has at least one zero row or zero column. Thus, $\mathrm{\det}(\mathbf{A}_{i}-\mathbf{A}_{j}) = 0$ and $\mathrm{rank}(\mathbf{A}_{i}-\mathbf{A}_{j}) < L$. It remains to prove, by induction, that $\mathrm{rank}(\mathbf{A}_{i}-\mathbf{A}_{j})< L$ in the case that $\mathbf{A}_{i}-\mathbf{A}_{j}$ has exactly two non-zero entries in each row and each column.

When $L=2$, there are only $2!$ permutation matrices to be considered. Obviously, $\mathrm{rank}(\mathbf{A}_{i}-\mathbf{A}_{j})< 2$. Assume that when $L= m$, $\mathrm{rank}(\mathbf{A}_{i}-\mathbf{A}_{j})< m$. When $L=m+1$, assume that the ($i,1$) and ($j,1$) entries are $1$ in the first column and then add the entire $i^{th}$ row to the $j^{th}$ row in $\mathbf{A}_{i}-\mathbf{A}_{j}$. Remove the row and column where ($i,1$) entry locates and form a new matrix of size $\mathbf{M}$ of size $m$. Note that $\det(\mathbf{A}_{i}-\mathbf{A}_{j}) = \det(\mathbf{M})$. In addition, the $j^{th}$ row in $\mathbf{M}$ either has all zero entries or contains exactly two non-zero entries. In the former case, $\det(\mathbf{M}) = 0$. In the latter case, $\mathbf{M}$ has exactly two non-zero entries in each column and each row. By induction assumption, $\mathrm{rank}(\mathbf{M}) < m$, and hence $\det(\mathbf{M}) = 0$. We conclude that $\det(\mathbf{A}_{i}-\mathbf{A}_{j}) = 0$ and (\ref{eqn:combination_network_vector_LNC_1}) does not hold for any $L$. This completes the proof.

\subsection{List of Notation}
\label{List_of_Notation}
\begin{longtable}{ll}
$S$:     &     the set of source nodes.  \\
$T$:     &     the set of receivers.    \\
$S_t$   &      the subset of $S$ corresponding to receiver $t$. \\
$E$:       &   the set of unit-capacity edges, with a topological order assumed. \\
In($v$):   &   the set of incoming edges to node $v$.  \\
Out($v$):  &   the set of outgoing edges from node $v$.  \\
In($N$):   &   equal to $\bigcup_{v\in N}\mathrm{In}(v)$ for node set $N$. \\
Out($N$):  &   equal to $\bigcup_{v\in N}\mathrm{Out}(v)$ for node set $N$. \\
$\omega$: & the number of data units generated by $S$, equal to $|\mathrm{Out}(S)|$.   \\
$\omega_t$: & equal to $|\mathrm{Out}(S_t)|$.     \\
$\otimes$:  & the Kronecker product. \\
$\mathbf{K}_{d,e}$:  & the local encoding kernel for adjacent pair $(d,e)$, which is an $\omega L \times \omega L$ matrix, \\
& of an $(L', L)$-fractional linear code. \\

$\mathbf{F}_{e}$:  & the global encoding kernel for edge $e$, which is an $\omega L \times L$ matrix, of an $(L', L)$-\\
& fractional linear code. \\

$\mathbf{G}_{s}$:  & the $|\mathrm{Out}(s)|L' \times |\mathrm{Out}(s)|L$ encoding matrix at source $s$ of an $(L', L)$-fractional \\
& linear code. \\

$\mathbf{m}_{e}$:  & the data unit transmitted on edge $e$. \\
$k_{d,e}$: & the local encoding kernel for adjacent pair $(d,e)$ of a scalar linear code. \\
$\mathbf{f}_e$: & the global encoding kernel for edge $e$ of a scalar linear code. \\
$\mathbf{D}_{t}$:  & the decoding matrix at receiver $t$ of a linear solution.  \\
$[\mathbf{m}_{e}]_{e\in A}$: & the column-wise juxtaposition of $\mathbf{m}_e$ with $e$ orderly chosen from subset $A$ of $E$. \\

$[\mathbf{K}_{d,e}]_{d,e \in A}$: & the block matrix consisting of $\mathbf{K}_{d,e}$ with both the rows and the columns indexed \\
& by subset $A$ of $E$. \\

$\mathbf{I}_{n}$:  & the identity matrix of size $n$. \\
$\mathbf{C}_{L}$:  & the $L\times L$ cyclic permutation matrix defined in (\ref{eqn:cyclic_permutation_matrix}).\\
%$\mathbf{\Lambda}_{\alpha}$:  & xxx \\
$\mathcal{C}_\delta:$  & the set of circulant matrices defined in (\ref{eqn:set_C_delta}).
\end{longtable}

\section*{Acknowledgment}
The authors would like to appreciate the valuable suggestions by
the associate editor as well as anonymous reviewers to help
improve the quality of the paper.

\end{document}